\documentclass[aps,prr,reprint,groupedaddress]{revtex4-2}
\usepackage{graphicx}
\usepackage{xcolor}
\usepackage[colorlinks=true,urlcolor=blue,citecolor=blue,linkcolor=blue]{hyperref}
\usepackage{gensymb}

\usepackage{physics}
\usepackage{siunitx}
\usepackage{xspace}
\usepackage{bm}
\usepackage{comment}
\usepackage{MnSymbol}

\newcommand{\bk}{\textbf{k}}

\newcommand{\bb}{\textbf{b}}
\newcommand{\bkk}{\textbf{K}}
\newcommand{\bq}{\textbf{q}}
\newcommand{\bqq}{\textbf{Q}}
\newcommand{\bgg}{\textbf{G}}

\newcommand{\br}{\textbf{r}}
\newcommand{\bp}{\textbf{p}}
\newcommand{\brr}{\textbf{R}}
\newcommand{\moire}{moir\'e\xspace}
\newcommand{\Moire}{Moir\'e\xspace}

\newcommand{\bvv}{\textbf{V}}
\newcommand{\btt}{\textbf{T}}
\newcommand{\bpi}{\mathbf{\Pi}}

\usepackage[normalem]{ulem}

\definecolor{SD-color}{named}{green}
\definecolor{AJ-color}{named}{blue}
\definecolor{GB-color}{RGB}{128,0,128}

\definecolor{SD-color2}{named}{green}
\definecolor{AJ-color2}{named}{blue}

\definecolor{GB-color2}{named}{red}

\begin{document}


\title{Exciton interacting with a \moire lattice\textemdash polarons, strings, and quantum probing of spin correlations}




\newcommand{\affiliationAarhus}{Center for Complex Quantum Systems, Department of Physics and Astronomy, Aarhus University, Ny Munkegade, DK-8000 Aarhus C, Denmark}

\author{Aleksi Julku}
\affiliation{\affiliationAarhus}
\author{Shanshan Ding}
\affiliation{\affiliationAarhus}
\author{Georg M. Bruun}
\affiliation{\affiliationAarhus}

\date{\today}

\begin{abstract}
The ability to create and stack different atomically thin transition metal dichacogenide (TMD) layers on top of each other has opened up a   rich playground for exploring new and interesting two-dimensional (2D) quantum phases. As a consequence 
of this remarkable development, there is presently a need for 
 new sensors to probe these 2D layers, since conventional techniques for bulk materials such as X-ray and neutron scattering are inefficient. Here, we develop a general theory for how an exciton in a TMD monolayer couples to spin and charge correlations in an adjacent  \moire lattice created by a TMD bi-layer. Virtual tunneling of charge carriers, assumed for concreteness to be holes, 
 between the \moire lattice and the 
monolayer combined with the presence of bound hole-exciton states, i.e. trions, give rise to an effective interaction between the \moire holes and the exciton. In addition to the Umklapp scattering, we show that this interaction is 
spin-dependent and therefore couples the exciton to the spin correlations of the \moire holes, which may be in- as well as out-of-plane. We then use our theory to examine two specific examples where the \moire holes form in-plane ferromagnetic or anti-ferromagnetic order. In both cases, the exciton creates spin waves in the \moire lattice, which we analyse by using a self-consistent Born approximation that includes such processes to infinite order. We show that the competition between magnetic order and exciton motion leads to the formation of a well-defined quasiparticle  consisting of the exciton surrounded by a 
cloud of magnetic frustration in the  \moire lattice sites below. For the anti-ferromagnet, we furthermore demonstrate the presence of the  elusive geometric string excitations and discuss how they can be observed  via their smoking gun energy dependence on the spin-spin coupling, which can be tuned by varying the twist angle of the \moire bi-layer. 
All these phenomena have clear signatures in the exciton spectrum,
and as such our results  illustrate 
that excitons are promising quantum probes providing optical access to the spin correlations of new  phases  predicted to exist in TMD materials.
\end{abstract}


\maketitle


\section{Introduction} 
 Transition metal dichacogenides (TMDs) have emerged as a new and powerful platform 
 for exploring strongly correlated  physics in two dimensions (2D). This is due to their rich spin-valley degrees of freedom and useful optical properties~\cite{Xiao2012,Cao2012,Zeng2012,Mak2012b,Schaibley2016,Wang2018},
 combined with the possibility to stack two or more monolayers with a  lattice mismatch or at a relative angle, which  creates a long wavelength \moire potential~\cite{Balents2020,Kennes2021,Andrei2021}. 
The  low energy physics of such \moire lattices 
 can be described by a highly tunable  Fermi-Hubbard model, 
 where many different many-body phases can be realised~\cite{Wu2018,Pan2020,Pan2020b,Pan2020c,Morales-Duran2022}. There  has recently been remarkable experimental progress exploring these systems including the observation of a Mott insulating state   
 at unit filling~\cite{Tang2020,Shimazaki2020} and a possible superconducting state~\cite{Wang2020}, correlated insulating states (generalised Wigner states) at discrete fractional fillings~\cite{Regan2020,Xu2020,Huang2021,Jin2021,Li2021c,Smolenski2021}, as well as the integer and fractional anomalous quantum Hall effects~\cite{Li2021b,Cai2023}. 
 Since excitons are tightly bound in TMDs, they can for many purposes be regarded as bosons, which together with the electrons 
 realise intriguing new Bose-Fermi mixtures. This has led to the observation of 
Fermi~\cite{Sidler2017} and Bose polarons~\cite{tan2022bose}, and has been predicted to  give rise to exciton 
mediated superconductivity~\cite{Laussy2010,Cotlet2016,Julku2022b, vonmilczewski2023superconductivity,zerba2023realizing}.
\Moire lattices are  naturally triangular and can 
 host a range of  magnetic and spin liquid phases~\cite{Hejazi2020,Pan2020b,Morales-Duran2022,Morales-Duran2023,Kiese2022} 
 with spin correlations predominantly in-plane~\cite{Zang2021,Zang2022,Kiese2022}. 
 Measurements of the out-of-plane magnetic susceptibility  using an external magnetic field
 indicate an antiferromagnetic  spin-spin coupling  in   WSe$_2$/WS$_2$~\cite{Tang2020} and MoTe$_2$/WSe$_2$~\cite{Li2021}  bi-layers at unit filling, whereas such measurements reveal ferromagnetic correlations away from the unit filling for a MoSe$_2$/WS$_2$ bi-layer~\cite{ciorciaro2023kinetic}.
 Transport measurements for a WSe$_2$/WSe$_2$ bi-layer are  consistent with a non-magnetic state such as a spin liquid~\cite{Ghiotto2021}.

In order to harness the full potential of these new 2D materials, it is important to have efficient probes for their properties. In particular, 
having  access to the  spin correlations of the \moire electrons  would be highly useful as they are often one of the defining features of the different magnetic and spin liquid states predicted to be the ground state in   \moire systems~\cite{Kiese2022,Zang2021,Pan2020b,Morales-Duran2022,Morales-Duran2023,Kiese2022,Drescher2022,Sherman2023,kuhlenkamp2022tunable}. Techniques developed for 3D materials such as  X-ray and neutron scattering are inefficient for 2D materials  due to the required large sample mass to obtain observable signals~\cite{Calder2019}, and it is difficult to extract the underlying particle correlations from magneto-transport or scanning probes~\cite{Papaj2022}.

Inspired by these  developments,  we explore in this work the properties of an exciton in a TMD monolayer 
placed on top of a TMD bi-layer forming a 
\moire lattice. We show that virtual tunneling of holes, which are 
taken to be the charge carriers, between the \moire lattice and the monolayer gives rise to a 
strong and spin-dependent exciton-hole interaction. This is because only holes and excitons with opposite spin interact significantly and can bind to form a trion. 
We show that this interaction gives rise to  Umklapp scattering to first order of the number of exciton-hole scattering events, whereas higher order 
scattering terms 
couple the exciton to the spin-spin correlations of the \moire holes. Our theory is then applied to two  cases, where \moire 
holes either form anti-ferromagnetic (AFM) or ferromagnetic (FM) order. 
The exciton is demonstrated to excite spin waves in both the cases, and to study the exciton spectral function, we employ a strong coupling theory based on a self-consistent Born approximation
that includes spin waves to infinite order. We find that these spin wave excitations result in the formation of  a well-defined quasi-particle  consisting of the exciton dressed by a cloud of magnetic frustration in the \moire lattice sites below. 
In the case of the AFM, there are also damped excited states, which  can be identified as geometric strings due to their characteristic energy dependence on the spin-spin coupling strength in the \moire lattice. Importantly, this dependence can be experimentally probed by changing the twist angle of the \moire bi-layer showing how the flexibility of the TMD setup can be used to detect these elusive states. Since our theory for the interaction between an exciton and the spin and charge degrees of freedom in an adjacent  \moire lattice is quite general, our results demonstrate how excitons can be used as non-evasive 
quantum probes for the new and rapidly growing class of 2D van der Waals materials.

The next sections of our paper are organised as follows. In Sec.~\ref{setup_sec}, we describe our setup and the resulting Hamiltonian. Section \ref{ScatteringSec}
analyses the exciton-hole scattering and discusses qualitatively 
the main physical effects arising from this. Next, we present in Sec.~\ref{SE_sec} a theory, which expands   
 the exciton energy spectrum in terms with an increasing number of exciton-hole scattering events, and we show how the exciton couples to the spin-spin correlations 
 of the \moire holes. In 
 Sec.~\ref{AFM_FM_SCBA_sec}, we  combine this  theory with a non-perturbative   Born approximation to describe the  
 cases where the \moire holes form in-plane AFM or FM order. This leads to a physical picture where the exciton 
 creates and absorbes spin waves. 
 Our numerical results including the emergence of polarons, damping, and geometric strings are presented in Sec.~\ref{NumSec}, and we end in Sec.~\ref{ConcSec} with a discussion 
 and outlook.

\begin{figure}[ht!]
\centering
\includegraphics[width=1.0\columnwidth]{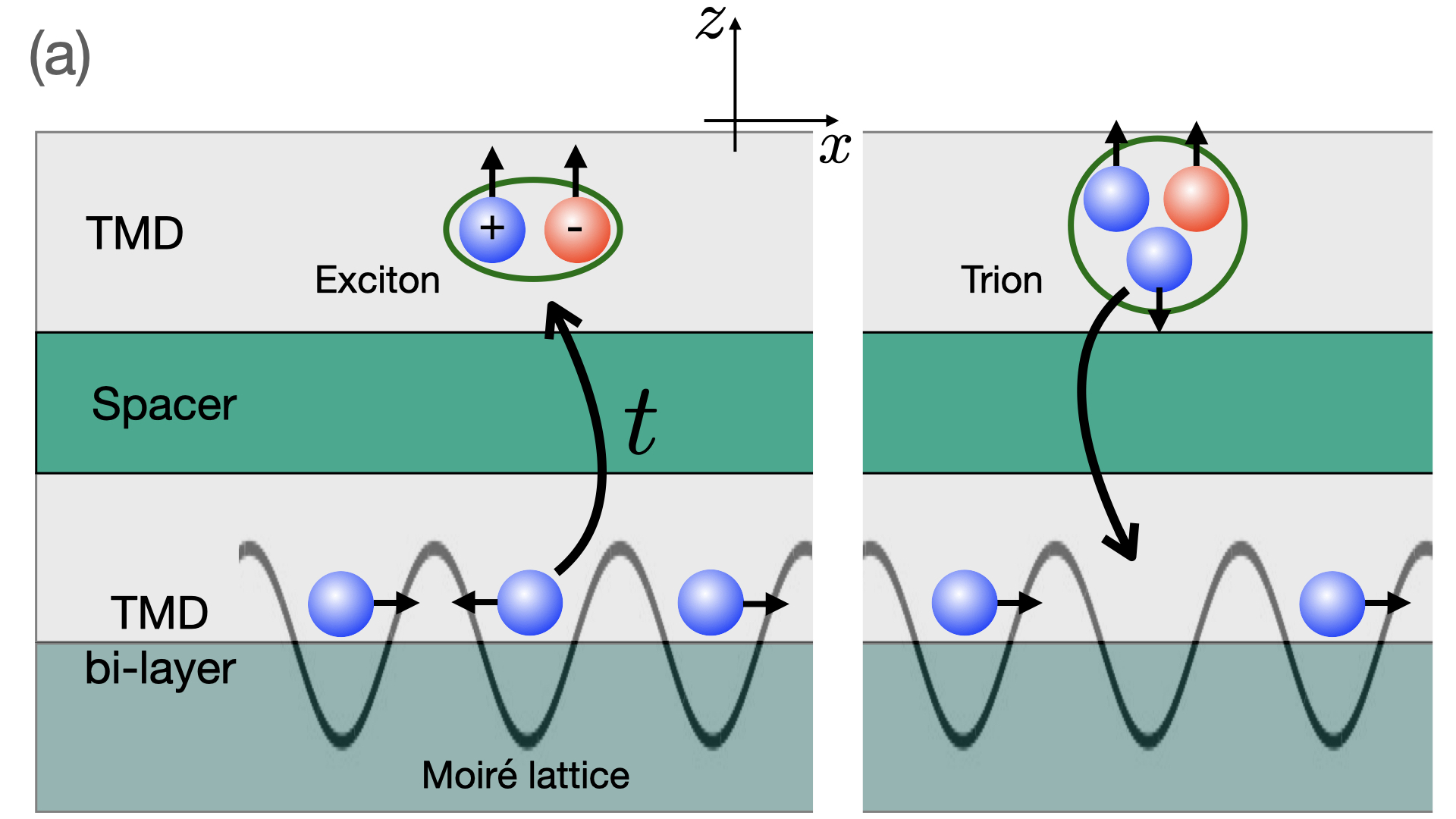}

\vspace{0.2cm}

\includegraphics[width=1.0\columnwidth]{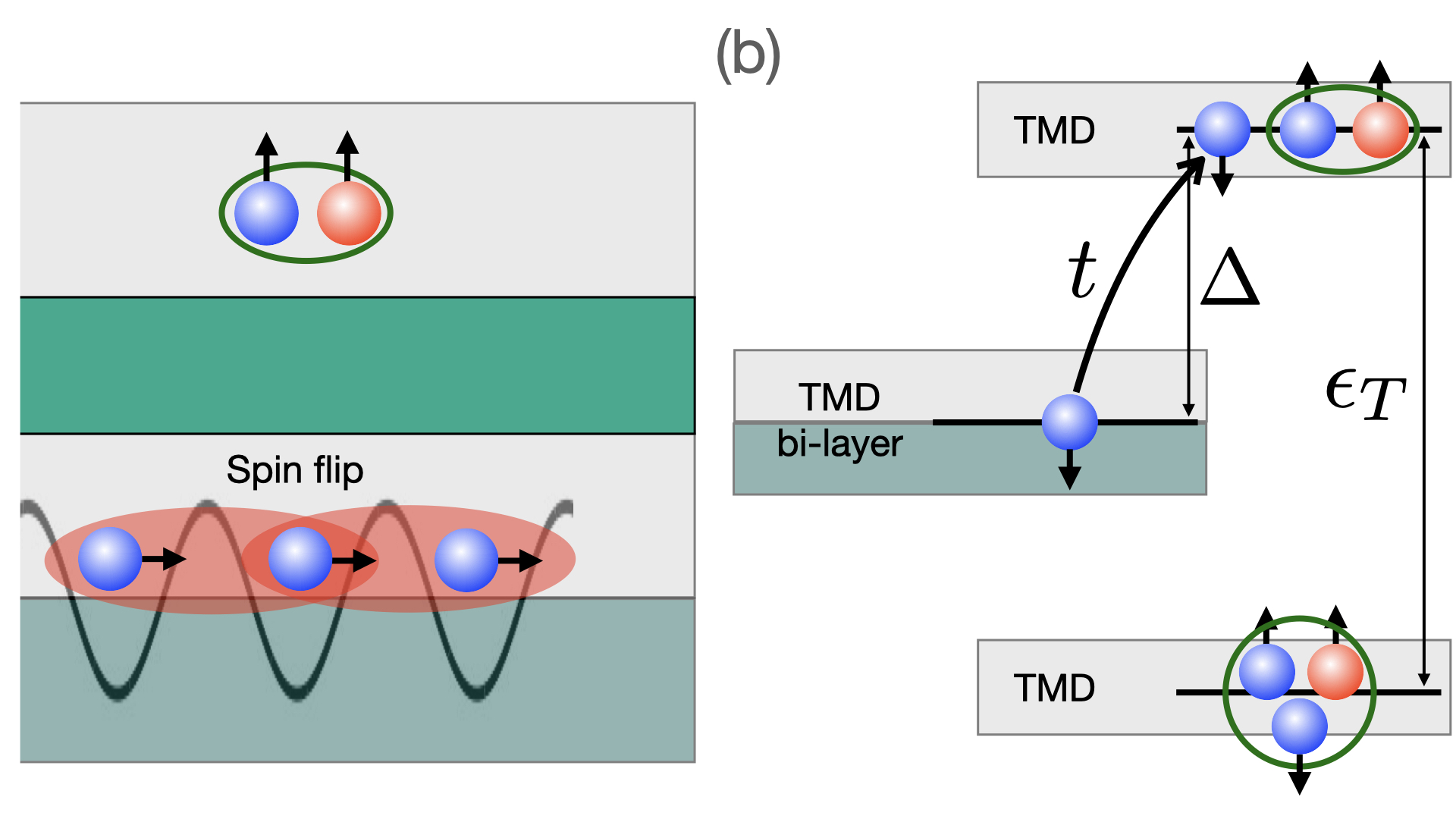}
\caption{(a) System considered. A top TMD layer contains an exciton formed by a  spin $\uparrow$ hole (blue ball) 
bound to a spin $\uparrow$ electron (red ball). An adjacent TMD bilayer forms a \moire lattice with one hole per site featuring in-plane spin correlations, which in the figure are taken to form an 
AFM. A spin $\downarrow$ \moire hole can virtually tunnel to the top layer and form a trion state with the exciton. Since its spin in the trion state is perpendicular to the TMD layers, 
this can lead to a flip of its in-plane spin when it tunnels back to the \moire lattice.  This tunnelling gives rise to an effective exciton-hole interaction that creates magnetic frustration (red ellipses) in the \moire lattice sites below the exciton. (b) Underlying level structure. A spin $\downarrow$ hole in the TMD bi-layer forming the \moire lattice can tunnel to the top TMD layer where its energy is increased by $\Delta$. It can however bind with an exciton thereby lowering its energy by the trion binding energy $\epsilon_T$.   
}
\label{Fig1r}
\end{figure}

\section{Setup}\label{setup_sec}
We consider the setup illustrated in Fig.~\ref{Fig1r}(a) where a TMD mono-layer is stacked on top of a TMD bi-layer forming a \moire lattice. 
We will explore how the spin correlations in the \moire lattice
affect an exciton in the mono-layer, which will be refered to as the exciton layer in the following.
To avoid the formation of a three-layer \moire system, the probe and \moire layers are separated by a 
spacing layer with energy  bands highly detuned by an energy off-set $\tilde \Delta$ with respect to the relevant energies.

For concreteness, we assume that the \moire lattice is in the hole doping regime; the analysis for the electron-doping case is completely analogous. Using the  Schrieffer-Wolff transformation, we obtain that the tunneling rate of holes between the top layer of the \moire system and the exciton layer is 
 $t\sim {\tilde{t}}^2/\tilde\Delta$, where $\tilde t \ll \tilde \Delta$
 is the interlayer tunneling between the spacer layer and its two neighbouring layers. Further details are given in App.~\ref{tunn_sec}. 

\subsection{Exciton coupled to \moire holes}
Without losing generality, we consider a spin-up ($\uparrow$) exciton in the exciton layer. Note that due to the spin-orbit coupling in the TMDs, 
 the spin quantisation axis is perpendicular to the layers, which we define as the $z$-axis, see Fig.~\ref{Fig1r}. There are at least two physical effects giving rise to a spin-dependent exciton-hole interaction. First, a $\uparrow$ hole tunneling from the \moire system to the exciton layer can be exchanged with the $\uparrow$ hole bound in the exciton whereas a $\downarrow$ hole cannot. This gives rise to a spin-dependent  hole-exciton exchange interaction, which however is quite weak.
   
  We  therefore focus on a second effect arising from the fact that the $\uparrow$ exciton interacts predominantly with
   $\downarrow$ holes. This is because excitons in TMDs are tightly bound and have a small spatial size. The Pauli exclusion principle, which prohibits two holes with the same spin to be at the same position therefore suppresses the 
   probability that a spin $\uparrow$ hole is close to a $\uparrow$ exciton and hence their interaction.
The effective Hamiltonian describing the coupling between the \moire system and the exciton layer is 
\begin{gather}
\label{FeshbachHam}
H =\sum_{\bk}
\begin{bmatrix}
a_{\bk\downarrow}^\dagger& h_{\bk\downarrow}^\dagger
\end{bmatrix}
\begin{bmatrix}
\epsilon_{\bk}+\Delta&t\\t&\epsilon_{\bk}
\end{bmatrix}
\begin{bmatrix} 
a_{\bk\downarrow}\\ h_{\bk\downarrow}
\end{bmatrix}+\nonumber\\
\sum_{\bk} \epsilon^x_{\bk}x_{\bk}^\dagger x_{\bk}+ \sum_{\bk\bk'\bq}V(\bq)x_{\bk+\bq}^\dagger a_{\bk'-\bq\downarrow}^\dagger a_{\bk'\downarrow}x_{\bk}.
\label{Hamiltonian}
\end{gather}
Here $a_{\bk\downarrow}^\dagger$ and $h_{\bk\downarrow}^\dagger$ create a $\downarrow$ hole with 
in-plane momentum $\bk$ and kinetic energy $\epsilon_{\bk}$ in the  exciton layer
and top layer of the \moire system respectively, $x_{\bk}^\dagger$ creates an exciton in the  exciton layer with momentum $\bk$ and kinetic energy $\epsilon^x_{\bk}=k^2/2m_x$ ($m_x$ being the exciton mass),
 $\Delta$ is the energy off-set between the exciton layer and the \moire system,
  and $V(\bq)$ is the interaction between the exciton and the $\downarrow$ holes in the exciton layer. We
   have ignored the tunneling of $\uparrow$ holes as they are assumed to interact only weakly with the excitons, 
   and we use units where $\hbar=1$.

\subsection{\Moire bi-layer}\label{Moiresystem}
\begin{figure}[ht!]
\centering
\includegraphics[width=0.8\columnwidth]{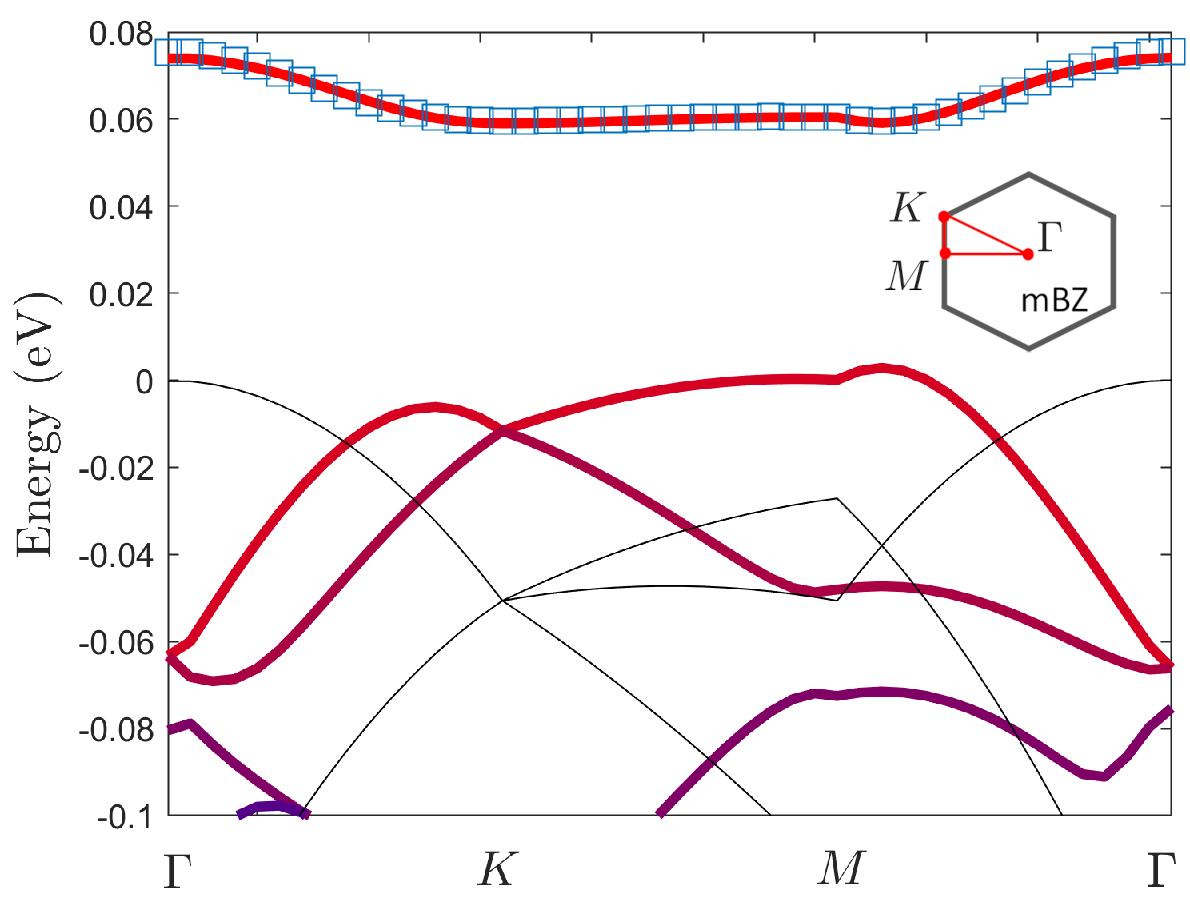}%
\caption{Energy dispersions (red lines) of the highest spin $\downarrow$ valence Bloch bands of the MoSe$_2$-WS$_2$ bilayer along high symmetry directions in the mBZ for twisting angle 
$\theta=2.5^\circ$. The black lines are the Bloch bands 
 without the \moire coupling. Squares depict the dispersion of the highest valence band obtained from the nearest-neighbour tight-binding model.
 }
\label{Fig_disp}
\end{figure}

In addition to Eq.~\eqref{Hamiltonian} describing the exciton and its coupling to the \moire holes, we also need a Hamiltonian for the \moire system itself.
While our results are quite general, we focus in the rest of the paper on the experimentally relevant case of a half-filled triangular \moire lattice formed by a MoSe$_2$-WS$_2$ bilayer.
To describe  the \moire lattice, we use a microscopic continuum model with interlayer hole tunneling between the MoSe$_2$  and WS$_2$ layers~\cite{Ruiz-Tijerina2019,Alexeev2019}. For brevity, we present the details and numerical parameters of these calculations in App.~\ref{moiresec} and simply plot in Fig.~\ref{Fig_disp} the resulting highest valence Bloch bands along high symmetry directions in the \moire Brillouin zone (mBZ) for a twist angle $\theta = 2.5^\circ$ between the two layers. We see that the highest \moire valence band is well separated in energy  from the other bands. 
As we are interested in the hole-doping regime of the half-filled highest band, we discard the lower bands and describe the \moire holes in the highest valence band with an effective triangular tight-binding  model. By accounting also for the repulsive Coulomb interaction between the \moire holes, we can write down an effective many-body \moire Hamiltonian for the \moire holes as
\begin{align}
\label{Hm}
&H_m = \sum_{ ij  \sigma} t^\sigma_{ij} h^\dag_{i\sigma} h_{j\sigma} + \sum_i U_0 h^\dag_{i\uparrow}h_{i\downarrow}^\dag h_{i\downarrow} h_{i\uparrow} \nonumber \\
&+ \sum_{ij\sigma\sigma'} U_{ij}  h^\dag_{i\sigma}h_{j\sigma'}^\dag h_{j\sigma'} h_{i\sigma} + \sum_{ij \sigma \sigma'} X_{ij}  h^\dag_{i\sigma}h_{j\sigma'}^\dag h_{i\sigma'} h_{j\sigma},     
\end{align}
where $h_{i\sigma}$ annihilates a hole with  spin  $\sigma \in \{\uparrow,\downarrow\}$  at  \moire lattice site $i$. The first term in Eq.~\eqref{Hm} describes hole hopping, where the matrix elements $t_{ij}^\sigma$ are  obtained by taking the Fourier transform of the energy dispersion of 
the highest \moire valence band. For small twisting angle $\theta$, corresponding to a large \moire lattice constant, we find as expected that the nearest-neighbor 
hopping matrix element is by far  the largest in magnitude. In Fig.~\ref{Fig_disp} the band dispersion produced by the first term of Eq.~\eqref{Hm} with only nearest neighbour hopping (squares)  is compared to the original highest valence band. We see that the agreement is quantitative, which 
justifies the use of the effective single band model given by Eq.~\eqref{Hm}. 

The second, third and fourth terms in Eq.~\eqref{Hm} describe, respectively, the on-site, non-local direct and exchange  Coulomb interaction  between the \moire holes,
where the interaction matrix elements $U_0$, $U_{ij}$ and $X_{ij}$ are  
$U_0 = \langle w_i w_i | V_c | w_i w_i \rangle$, $U_{ij} = \langle w_i w_j | V_c | w_i w_j \rangle$, 
and $X_{ij} = \langle w_i w_j | V_c | w_j w_i \rangle$.
Here $V_c(r)$  is the Coulomb interaction and  $w_i$ is the Wannier function for the \moire lattice site $i$~\cite{Morales-Duran2022,Julku2022}. Note that these Wannier states have components in both layers of the \moire bilayer leading to 
intra- and inter-layer Coulomb matrix elements. A detailed account of these calculations is given in App.~\ref{moiresec}.

As expected, we find that for small twist angles the on-site interaction $U_0$ is much larger than any other relevant energy scale. For example, for $\theta = 2.5^\circ$ we have $U_0/U_{\text{NN}} \sim 4$ and $U_0/|t_{\text{NN}}^\sigma| \sim 25\gg 1$ where "NN" refers to the nearest-neighbour terms. It follows that doubly
occupied sites are energetically suppressed and that the 
 low energy excitations of the holes at half filling can be described by an extended Heisenberg model 
\begin{equation}
H_m=J_1\sum_{\langle i,j\rangle}{\mathbf s}_i\cdot{\mathbf s}_j+J_2\sum_{\llangle i,j\rrangle}{\mathbf s}_i\cdot{\mathbf s}_j + H_{\text{DM}}.
\label{SpinH}
\end{equation}
Here,  the spin operators read 
 ${\mathbf s}_i=\sum_{\sigma\sigma'}h_{i\sigma}^\dagger\boldsymbol{\sigma}_{\sigma\sigma'}h_{i\sigma'}/2$
where $\boldsymbol{\sigma}=(\sigma_x,\sigma_y,\sigma_z)$ is a vector of  the Pauli matrices. Microscopic expressions for the nearest neighbour  $J_1$ and next-nearest neighbour $J_2$
spin-spin coupling strengths are derived in Ref.~\cite{Morales-Duran2022} and for completeness given in App.~\ref{moiresec}.
Using the experimentally realistic parameters for the  MoSe$_2$-WS$_2$ bilayer resulting in the 
Bloch bands shown in Fig.~\ref{Fig_disp}, 
  we obtain $J_1 \sim 0.23$meV ($J_1\sim 0.95$ meV) and $J_2 \sim 0.001$meV ($J_2\sim 0.02$ meV) for the twisting angles $\theta=2.5^\circ$ ($\theta=3.3^\circ$).

Our calculations also yield a small imaginary part for the NN hopping matrix elements $t^\sigma_{\text{NN}}$. This  
 gives a rise to the so-called Dzyaloshinskii-Moriya term $H_{{\text{DM}}} = J_1\sum_{\langle i,j \rangle}[(\cos 2\phi-1) \mathbf{s}_i \cdot \mathbf{s}_j + \sin 2\phi \hat{e}_z \cdot (\mathbf{s}_i \cross \mathbf{s}_j )]$, where $\phi$ is the complex phase of $t_{\text{NN}}$~\cite{Zang2021,Kiese2022}. While this term is small for small $\phi$ and therefore omitted  in the subsequent sections, its presence 
 is important since it breaks the $O(3)$ symmetry of the \moire lattice and has been predicted to favour \emph{in-plane} 
magnetic ordering~\cite{Zang2021,Zang2022,Kiese2022}.

Note that it is a very challenging to provide a microscopic description of the highly correlated states of electrons in  triangular \moire lattices
and the microscopic model used may not be quantitatively accurate. This is however not essential for the main results of the present paper; namely  
that the coupling between an exciton and the in-plane spin correlations of electrons (holes) in a \moire lattice leads to the formation of  polarons and geometric strings, which 
are visible in the exciton spectrum making it a useful quantum probe. Indeed, different values of the spin-spin 
couplings will of course change our results quantitatively but not qualitatively as long as the underlying phases remain 
the same.

\section{Exciton-hole scattering}\label{ScatteringSec}
 While the interaction $V(\bq)$ in Eq.~\eqref{Hamiltonian} takes place in the exciton layer between spin $\downarrow$ holes and a spin $\uparrow$ exciton, the tunneling $t$ gives rise to  an effective scattering between a spin $\downarrow$ hole in the top \moire layer and an exciton in the exciton layer~\cite{Schwartz2021,Kuhlenkamp2022}. Indeed, a spin $\downarrow$ hole in the top \moire layer can tunnel to the exciton layer, interact with the exciton, and subsequently tunnel back to the top \moire layer.

\begin{figure}[ht!]
\centering
\includegraphics[width=1\columnwidth]{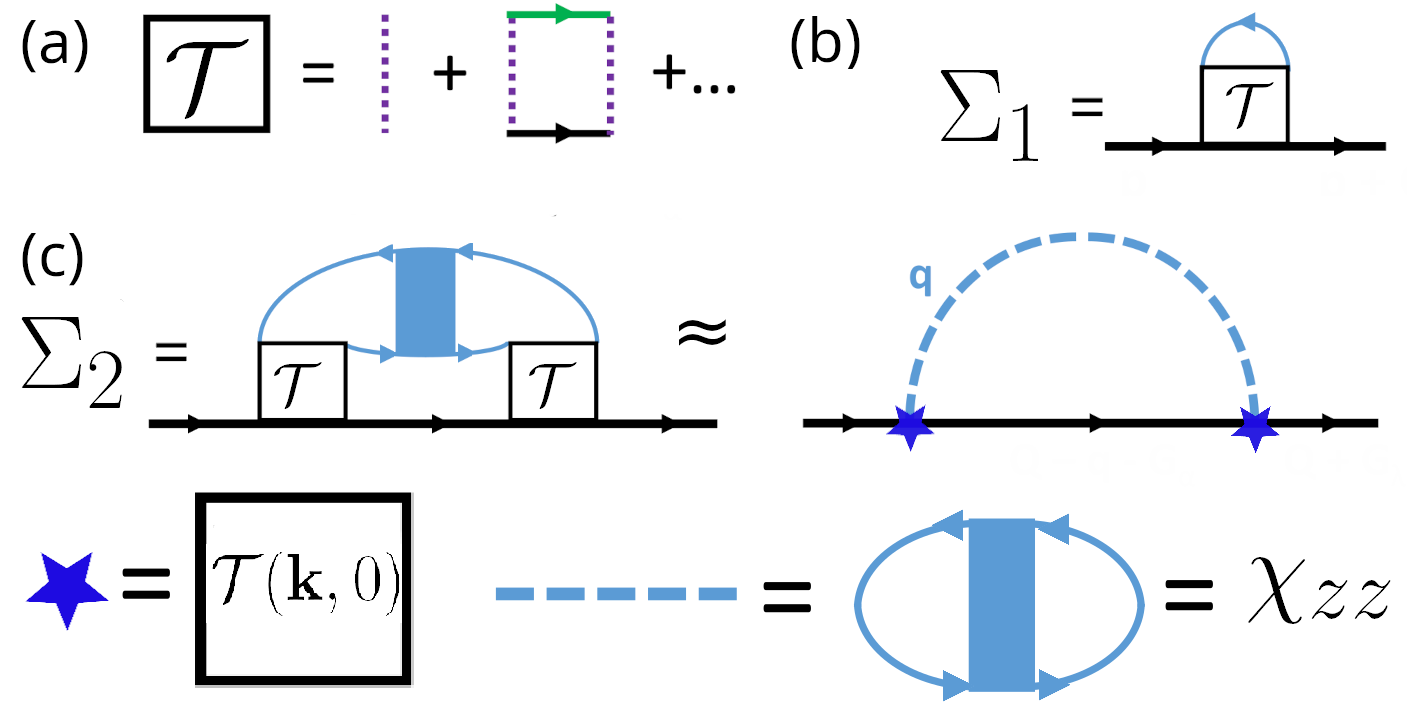}%
\caption{(a) The scattering matrix for an exciton (black line) and a spin $\downarrow$ hole in the exciton layer (green line) in the ladder approximation, with the dashed purple lines depicting the exciton-hole interaction $V(\bq)$. (b) The exciton self-energy term $\Sigma_1$. The blue line corresponds to a \moire hole (c) The exciton self-energy term $\Sigma_2$ with the blue box indicating correlations between a spin $\uparrow$ and a spin $\downarrow$ hole in the \moire lattice. The star indicates 
the static scattering matrix and the dashed blue line a spin wave in the \moire system. (d) The self-consistent Born approximation for $\Sigma_2$. The double line is the full exciton Green's function. 
}
\label{Fig2}
\end{figure}

By taking such tunneling processes into account to the lowest order in $t/\Delta$, we show in App.~\ref{Tmat_sec} that the scattering matrix between spin $\downarrow$ holes 
in the top \moire layer and a spin $\uparrow$ exciton in the exciton layer is given by $t^2{\mathcal T}/\Delta^2$.  
Here, ${\mathcal T}$ is the scattering matrix between a hole and an exciton in the exciton layer, which can be evaluated within the ladder approximation as depicted diagrammatically in Fig.~\ref{Fig2}(a). Since one can to a good approximation  ignore the momentum dependence of the interaction 
$V(\bq)$~\cite{Efimkin2020}, we obtain
\begin{equation}
\label{Tmatrix}
{\mathcal T}({\mathbf K},\omega)=\frac{1}{\Pi(0,\epsilon_T)-\Pi({\mathbf K},\omega)},
\end{equation}
where
 $\Pi({\mathbf K},\omega)$ is the  propagator of an exciton-hole pair in the exciton layer with total momentum and energy $({\mathbf K},\omega)$.
 In Eq.~\eqref{Tmatrix}, we have replaced the bare interaction strength $V(\bq)$ with the binding energy $\epsilon_T<0$ of a trion state consisting of 
a spin $\uparrow$ electron, a spin $\uparrow$ hole, and a spin $\downarrow$ hole in the exciton layer
as explained in more detail in App.~\ref{Tmat_sec}~\cite{Wouters2007,Carusotto2010,Bastarrachea-Magnani2019}.

Equation \eqref{Tmatrix} takes into account that the interaction $V(\bq)$ can support a bound state between 
 a $\downarrow$ hole in the exciton layer and a $\uparrow$ exciton, i.e.\ a trion as illustrated in Fig~\ref{Fig1r}(a). 
 This gives rise to a pole in the scattering matrix at the trion energy, and in order to simplify the subsequent numerical 
 calculations, we expand around this pole  writing  ${\mathcal T}(\bk,\omega) =Z/[\omega - \epsilon_T(\bk)]$. 
 Here, $\epsilon_T(\bk) = \Delta + \epsilon_T + k^2/2m_T$ is the energy of the trion with momentum $\bk$ and 
residue $Z$ with $m_T$ being the trion mass, 
see Fig.~\ref{Fig1r}(b) and App.~\ref{Tmat_sec}. As discussed further in  Sec.~\ref{NumSec}, the trion energy $\epsilon_T$ is in typical experiments 
well below any relevant energies~\cite{Tan2020}, and we can consequently to a good approximation neglect the frequency dependence of the scattering matrix   writing ${\mathcal T}({\mathbf p})\equiv{\mathcal T}({\mathbf p},\omega=0)$.

\subsection{Effective interaction}\label{phys_sec}
We first discuss heuristically  the main physical consequences  of the 
 exciton-hole scattering given by Eq.~\eqref{Tmatrix}. In this way, we can qualitatively describe the main results presented in this paper 
 before getting  into the  rigorous details of our calculations.

\begin{figure}[ht!]
\centering
\includegraphics[width=1.0\columnwidth]{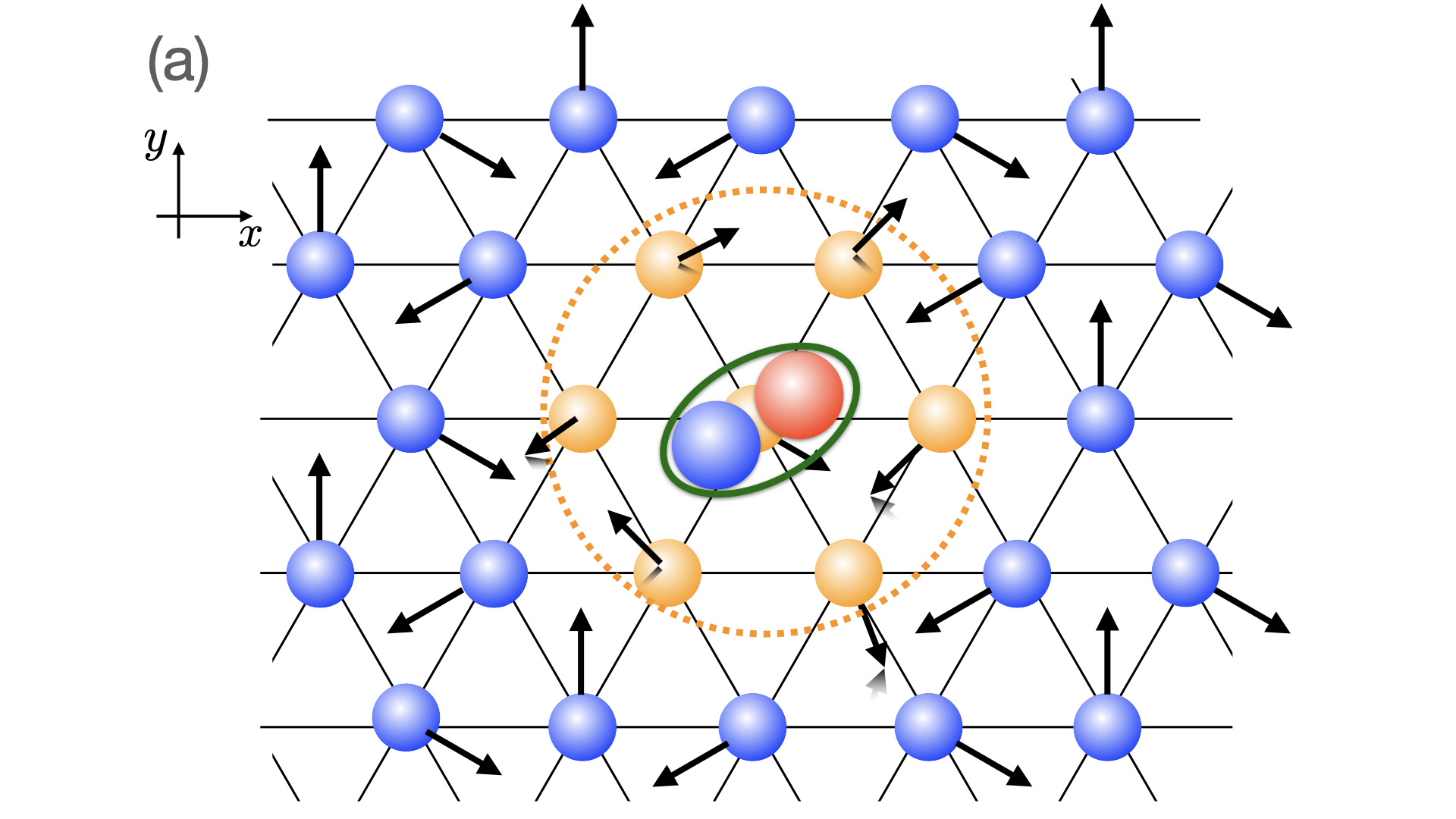}

\vspace{0.2cm}

\includegraphics[width=0.85\columnwidth]{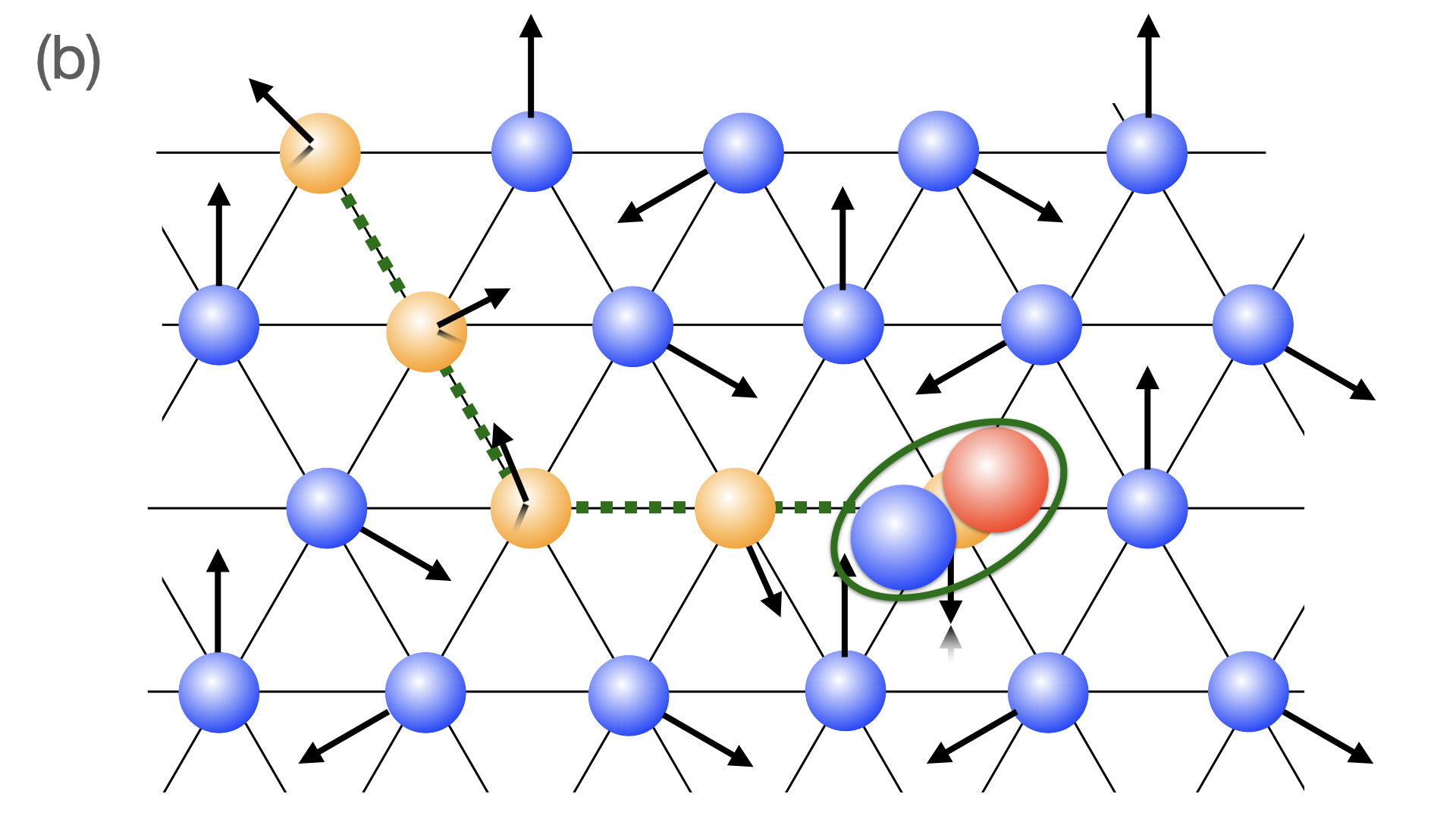}
\caption{(a) The exciton-hole interaction Eq.~\eqref{EffectiveInt} flips spins in a $120\degree$ AFM formed by the \moire holes, which leads to the formation of a quasiparticle consisting of an exciton in the top TMD layer and holes with miss-aligned spins (orange balls) in the \moire lattice below. (b) As the exciton moves, it leaves a trace (green dashed line) of mis-aligned spins  in the \moire lattice below. This creates a linear potential for the exciton, which supports string-like excitations.   
}
\label{Fig1}
\end{figure}

 While the scattering matrix given by Eq.~\eqref{Tmatrix} is written in the  momentum space, it is illuminating to discuss the interaction in the
   real space. As  mentioned above, the \moire holes reside in a triangular \moire lattice described by the operators $h_{i\sigma}$. On the other hand, the parabolic  dispersion of the exciton corresponds to 
    a continuum system. However, as we will show explicitly in Sec.~\ref{UmklappSec}, scattering on the \moire holes gives rise 
     a triangular mean-field potential for the exciton with the same periodicity as the \moire lattice. We therefore for the sake of the present  discussion 
     introduce the operators $\gamma_i^\dagger$ creating an  exciton at site $i$  in a triangular lattice. Fourier transforming Eq.~\eqref{Tmatrix}  
     then yields the  effective exciton-hole interaction 
     $V_\text{eff} = \sum_{ij} \mathcal T(\br_i-\br_j) \gamma^\dag_i h^\dag_{i\downarrow} h_{j\downarrow} \gamma_j\simeq\sum_{i} \mathcal T(0) \gamma^\dag_i \gamma_i h_{i\downarrow}^\dag h_{i\downarrow}$, where $\mathcal T(\br)$ is the static exciton-hole scattering matrix in  real-space and we have used the fact that the dominant  term is the local one with $i=j$. Since we are interested in the case of a  half-filled \moire lattice, we have $h_{i\downarrow}^\dagger h_{i\downarrow}=1/2-s_{iz}$.
Using this gives 
\begin{equation}
V_\text{eff}\simeq  \sum_{i} \mathcal T(0) \gamma^\dag_i \gamma_i s_{iz}+V_\text{stat},
\label{EffectiveInt}
\end{equation}     
 where the last term is a  static potential term  not important for the present  discussion. Since we are  focusing on in-plane magnetic 
 order of the \moire holes perpendicular to the $z$-axis, 
 the $s_{iz}$ operators induce spin rotations as seen explicitly with the  identity $s_{iz} = {\exp(i \pi s_{iz}) }/{2i}$. It follows that Eq.~\eqref{EffectiveInt}
 describes an effective interaction where the exciton flips the in-plane spins of the  holes  below it in the \moire lattice as illustrated  in Fig.~\ref{Fig1}(a). 
 As discussed in Sec.~\ref{AFM} and App.~\ref{string_ex_sec}, a rigorous expression for the effective exciton-hole interaction 
 in real space  describes the same physics as  Eq.~\eqref{EffectiveInt} and differs mainly by including \moire Bloch wave functions.

As we shall discuss in detail below, this has two main consequences.  First, the exciton-hole interaction leads to the  creation of a \emph{quasi-particle} consisting of 
an exciton surrounded by  a cloud of magnetic frustration of the \moire holes in its vicinity as illustrated in Fig.~\ref{Fig1}(a). As we shall see, the energy and mass of this quasiparticle can differ 
significantly from those of the free exciton, which should be observable using spectroscopy. Second, since the exciton leaves a trace of magnetic 
frustration in its path as illustrated in Fig.~\ref{Fig1}(b), it experiences a linear potential with a slope proportional to $J_1$. This can trap the exciton resulting in the existence of excitations corresponding to \emph{geometric string states}, in analogy to what has been  predicted  for a hole moving in an 
AFM background~\cite{Liu1992,Trumper2004,Hamad2008,Manousakis2007}. As  we shall discuss in detail below, the \moire bi-layer offers new and promising ways to observe 
smoking gun features of these elusive states.

In the rest of the paper, we will present a  quantitative discussion of these effects.  

\section{Exciton self-energy}\label{SE_sec}
In this section, we set up a strong coupling theory describing how the charge and spin correlations in the \moire lattice  affect    
the exciton. Our diagrammatic approach is based on considering processes with an increasing number of exciton-hole scattering events. That is, we expand the exciton self-energy in increasing orders of the scattering matrix ${\mathcal T}$, keeping all first and second order diagrams.

\subsection{Umklapp scattering}\label{UmklappSec}
The exciton self-energy arising from a single exciton-hole scattering event is shown  diagrammatically in Fig.~\ref{Fig2}(b). It describes the creation of 
uncorrelated holes in the \moire lattice due to the scattering on the exciton and is given by  
\begin{align}
\Sigma_1({\mathbf p}',{\mathbf p})=&\frac{t^2}{2\Delta^2}\frac{1}{A}\sum_{\mathbf q}{\mathcal T}({\mathbf p}-{\mathbf q})u_\downarrow^*({\mathbf q})u_\downarrow(\bq + \bp' - \bp)\nonumber\\
&\times\Big(\frac{1}{2}\delta_{{\mathbf p}'-{\mathbf p},\bgg_\alpha}-\frac{1}{\sqrt N}\langle s^z_{\bp'-\bp}\rangle\Big).
\label{Selfenergy1}
\end{align}
Here, $A$ is the area of the system and $\langle\ldots\rangle$ denotes  the average with respect to the ground state of the \moire holes. Furthermore, $\bp$ ($\bp'$) is the incoming (out-going) exciton momentum, which is  conserved up to the reciprocal \moire lattice vectors $\bgg_\alpha$ since the exciton scatters on the $\downarrow$ holes residing in the 
lattice. 
In deriving Eq.~\eqref{Selfenergy1}, we have expanded the operator $h_{\bk \downarrow+{\mathbf G}_\alpha}^\dagger$, which creates a hole with momentum 
$\bk+{\mathbf G}_\alpha$ in the top layer of the \moire system, as
\begin{align}
\label{projection_main}
h_{\bk+{\mathbf G}_\alpha\downarrow}^\dagger&=\sum_nu_{n\downarrow}(-\bk-{\mathbf G}_\alpha)h_{n\bk \downarrow}\simeq u_{1 \downarrow}(-\bk-{\mathbf G}_\alpha)h^\dagger_{1\bk\downarrow}\nonumber\\
&=u_\downarrow(-\bk-{\mathbf G}_\alpha)\frac1{\sqrt{N}}\sum_je^{i\bk\cdot\br_j}h^\dagger_{i\downarrow}.
\end{align}
Here $\bk$ is a momentum in the mBZ, $h_{n\bk \downarrow}^\dagger$ creates a hole in \moire  Bloch band $n$ with momentum $\bk$ and spin $\downarrow$, and $u_{n\downarrow}({\mathbf k})$ is the projection of the corresponding Bloch function to the top \moire layer.  The approximation in the first line of Eq.~\eqref{projection_main} corresponds to 
 projecting to the highest \moire valence band ($n=1$),  since it is well separated from the other \moire bands as discussed in 
 Sec.~\ref{Moiresystem} and shown in Fig.~\ref{Fig_disp}. In the second line we have Fourier transformed to the real space with $N$ being the number of \moire sites, and suppressed the band index of the Bloch functions. Finally, we have in Eq.~\eqref{Selfenergy1}
 used the identity $h_{i\downarrow}^\dagger h_{i\downarrow}=1/2-s_{iz}$ valid at half-filling, and 
 defined  $s^z_{\mathbf p}=\sum_j\exp(-i{\mathbf p}\cdot {\mathbf r}_j)s_{j}^z/\sqrt{N}$. 
In App.~\ref{AppsecSigma1}, we give further details regarding the derivation of  Eq.~\eqref{Selfenergy1}.

The first term in Eq.~\eqref{Selfenergy1} describes Umklapp scattering of the exciton on holes residing in the \moire lattice in the Mott phase and it is present even when there is no magnetic order. The effects of the Umklapp scattering on the 
exciton spectrum have recently been used to probe  
 the formation of an electronic Wigner crystals and stripe phases at various filling fractions in \moire lattices~\cite{Shimazaki2020,Shimazaki2021,Xu2020,Jin2021,Miao2021,Smolenski2021}.   
 The  second term of Eq.~\eqref{Selfenergy1},  proportional to
$ \langle s^z_{\mathbf p}\rangle$, couples the exciton to any out-of-plane magnetic order of the \moire holes. 
It was recently shown that this term gives rise to observable effects on  
the exciton spectrum that can be used to probe such out-of-plane magnetic 
order~\cite{Salvador2022}. 

In the present work, we  instead wish to explore how the exciton couples to spin correlations in the \moire lattice, which can be in an \emph{arbitrary} direction.  
This is motivated by the fact that the Dzyaloshinskii-Moriya spin coupling  has been predicted to favor in-plane magnetic ordering~\cite{Zang2021,Zang2022,Kiese2022}, for which the Umklapp scattering term given by  
 $\Sigma_1$ is insensitive.

\subsection{Coupling to spin correlations}
We therefore need to analyse how the exciton couples to the spin correlations in the \moire lattice, which 
first occurs to second order in the  exciton-hole scattering. Such processes are described  by the self-energy term $\Sigma_2$ shown in 
Fig.~\ref{Fig2}(c), where the exciton creates an electron-hole pair in the \moire lattice, which is strongly correlated 
due to interactions in the \moire lattice proportional to $U$.  As detailed in App.~\ref{AppSecSigma2}, 
a long calculation gives
\begin{gather}
\Sigma_2({\mathbf p}',{\mathbf p},i\omega_n)=\frac{t^4}{\Delta^4}\frac{T}{A}\sum_{i\omega_m}\sum_{\mathbf q}{\mathcal G}({\mathbf p}-{\mathbf q},i\omega_n-i\omega_m)
\nonumber\\
\times g({\mathbf p},-{\mathbf q})g(\bp-\bq, \bp' - \bp +\bq) \chi_{zz}(\bq,i\omega_m) \delta_{\bp',\bp + \bgg_\alpha}.
\label{Selfenergy2}
\end{gather}
Here ${\mathcal G}({\mathbf q},i\omega_m)$ is the exciton Green's function, $\omega_n=2\pi n T$ ($n=0,\pm1,\ldots$)
are bosonic Matsubara frequencies  and $T$ is the temperature. We have defined the 
vertex function $g({\mathbf p},-{\mathbf q})=\sum_{\mathbf k}{\mathcal T}(\bp - \bk)u^*({\mathbf k})u({\mathbf k}+{\mathbf q})$ for an exciton with momentum $\bp$ exciting a particle-hole pair with total 
momentum $\bq$ in the \moire lattice.  

Importantly,   
$\chi_{zz}({\mathbf q},i\omega_m)$ is the Fourier transform of the correlation function $\chi_{zz}({\mathbf r}_i-{\mathbf r}_j,\tau)=\langle T_\tau s^z_i(\tau) s^z_j(0)\rangle$
where $\tau$ ($T_\tau$) is the imaginary time (time ordering operator. 
Equation \eqref{Selfenergy2} therefore explicitly demonstrates that the exciton is coupled to spin-spin correlations of the  \moire holes via the
second order exciton-hole scattering processes. Since these spin correlations can be in an arbitrary direction, i.e.\ in- as well as out-of-plane, 
this opens up a way to optically detect states with in-plane magnetic order and states with more subtle spin correlations by 
measuring the exciton spectrum.  Equation \eqref{Selfenergy2} is therefore  a main result of our paper.

\section{In-plane ferromagnetic and antiferromagnetic order}\label{AFM_FM_SCBA_sec}
Having developed a theory describing how the exciton couples to spin correlations of the holes in the \moire lattice, we now demonstrate 
 that this can strongly affect its properties. We  consider two concrete examples: one where the \moire holes form in-plane AFM order and one where they form  FM order. 

\begin{figure}[ht!]
\centering
\includegraphics[width=1\columnwidth]{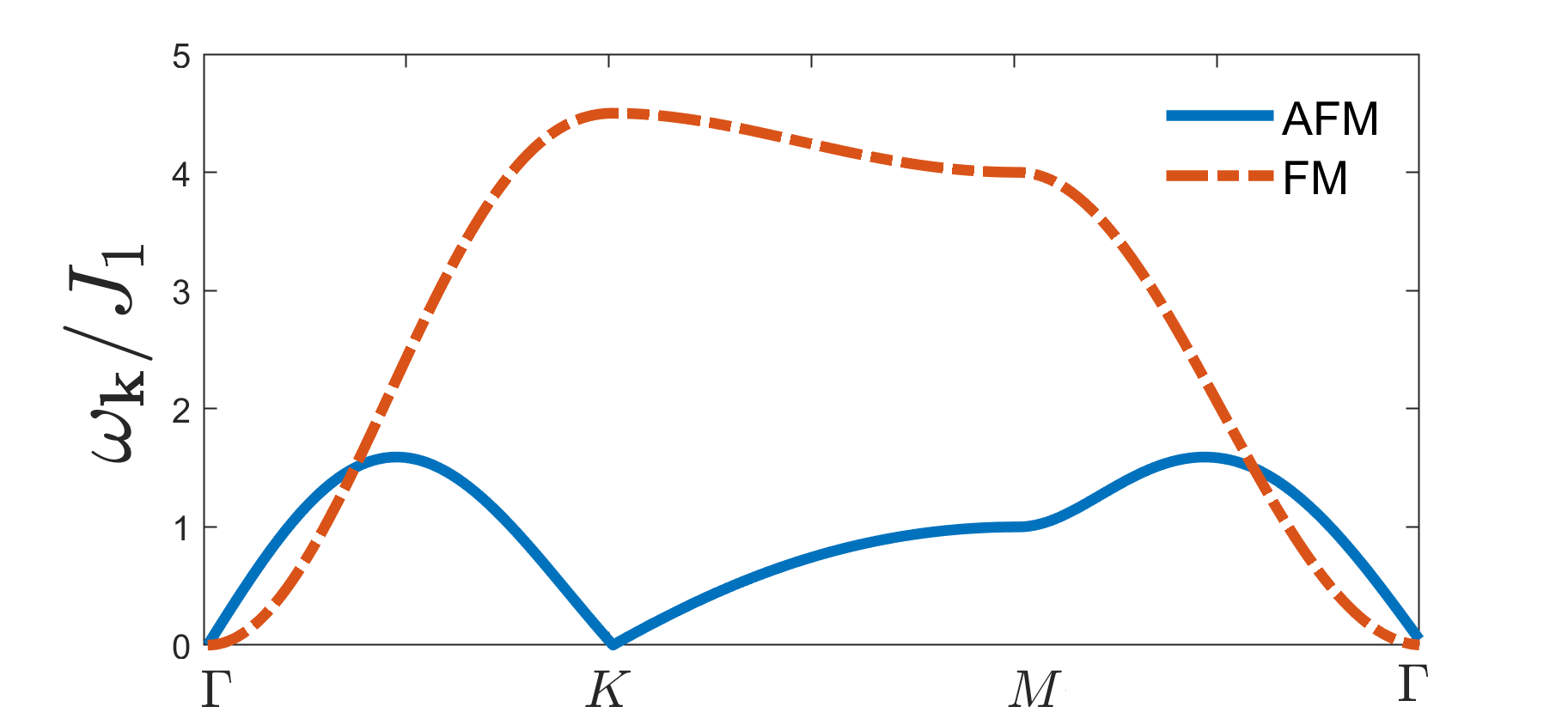}%
\caption{Spin wave spectrum of the \moire holes obtained from LSWT for the AFM (blue solid line) and FM (red dashed line) states as a function of momenta along high symmetry directions in the mBZ.
}
\label{Fig4}
\end{figure}

\subsection{Anti-ferromagnet}\label{AFM}
The values of $J_1$ and $J_2$ given in Sec.~\ref{Moiresystem} favor a $120\degree$  antiferromagnetic AFM ground state over other possible states such as spin liquids~\cite{Iqbal2016,Kiese2022,Drescher2022,Sherman2023}. 
By using linear spin wave theory (LSWT) to describe this $120\degree$ AFM phase, we can write Eq.~\eqref{SpinH} as 
\begin{equation}
H_m=\sum_{\mathbf k}\omega_{\mathbf k}b_{\mathbf k}^\dagger b_{\mathbf k},
\label{SpinWaveAFM}
\end{equation}
where $b_{\mathbf k}^\dagger$ creates a bosonic spin wave in the AFM  
 with crystal momentum ${\mathbf k}$ and energy $\omega_{\mathbf k}=3J_1\sqrt{(1-\gamma_{\mathbf k})(1+2\gamma_{\mathbf k})}/2$. 
Here $\gamma_{\mathbf k}=\sum_{\mathbf \delta}\cos({\mathbf k}\cdot{\boldsymbol{\delta}})/6$ is the structure factor with the sum taken over the six
nearest-neighbor links $\boldsymbol{\delta}$ in the triangular \moire  lattice~\cite{Kraats2022}, and  the small next-nearest coupling $J_2$ is neglected. We have in  Fig.~\ref{Fig4} plotted $\omega_\bk$   as a blue solid line  along the high-symmetry points of the mBZ. For small $k$ the spectrum is  isotropic and linear with  $\omega_{\bk}=ck$ where $c=3^{3/2}J_1a_m/4$  is the speed of the spin waves, and   $a_m$ is the \moire lattice constant. We note that linear spin wave theory accurately captures the low energy excitations of a triangular AFM in spite of its inherent geometric frustration~\cite{Chernyshev2009,Ferrari2019}.

Since the magnetic order is taken to be in-plane, i.e.\ perpendicular to the $z$-axis, we have $s^z_\bk = i(u_\bk - v_\bk) (b^\dag_{-\bk} - b_\bk)/2$
within LSWT, where $u_\bk = \sqrt{ \frac{2 + \gamma_\bk}{2 \sqrt{(1- \gamma_\bk)( 1 + 2\gamma_\bk)}} + \frac{1}{2}}$ and $v_\bk = \text{sign}(\gamma_\bk) \sqrt{ \frac{2 + \gamma_\bk}{2 \sqrt{(1- \gamma_\bk)( 1 + 2\gamma_\bk)}} - \frac{1}{2}}$ are  coherence factors. It follows that 
 $\chi_{zz}(\bk,\tau)$ is proportional to 
 $\langle T_\tau [b_{-\bk}^\dagger(\tau)-b_\bk(\tau) ][b_{-\bk}^\dagger(0)-b_\bk(0) ]\rangle$.
 Equation \eqref{Selfenergy2} therefore describes how the motion of the exciton creates and annihilates spin waves in the 
adjacent \moire AFM, since it couples differently to  spin $\uparrow$ and $\downarrow$ holes. This result is the quantitative momentum space 
version of the  heuristic real space arguments given in Sec.~\ref{phys_sec}. A precise real space version 
of Eq.~\eqref{Selfenergy2} is given in  App.~\ref{string_ex_sec} by Fourier 
 transforming the exciton-hole interaction to real space. It differs from the heuristic expression  
  Eq.~\eqref{EffectiveInt} mainly by including \moire Bloch wave functions but describes the same physics. 

The spin susceptibility is within the LSWT straightforwardly calculated to be  
\begin{equation}
\chi_{zz}(q) = -\frac{1}{4}\left[ \frac{(u_\bq - v_\bq)^2}{iq_n  - \omega_\bq} - \frac{(u_\bq - v_\bq)^2}{iq_n +\omega_\bq}\right]
\label{SzSz}
\end{equation}
where $iq_n$ is a bosonic Matsubara frequency. Technically, we have in this section used LSWT to express the strong particle-hole correlations in the \moire lattice in terms of low-energy spin waves, which is illustrated diagrammatically in Fig.~\ref{Fig2}(c).

\subsection{Ferromagnet}
As has been shown~\cite{Morales-Duran2022}, \moire bi-layers can, depending on the system parameters such as the twist angle, specific TMD materials,
and the dielectric constant of the surrounding medium, also feature negative values for $J_1$, which imply that the holes form a ferromagnet at half filling. 
We therefore also examine the case of an in-plane FM order assuming a negative value of $J_1$ in \eqref{SpinH}. In this case, the LSWT Hamiltonian is still given by  
Eq.~\eqref{SpinWaveAFM} but with  the spectrum 
 $\omega_\bk = 3J_1( \gamma_\bk -1)$, see red dashed line in Fig.~\ref{Fig4}. 
This excitation spectrum is gapless at $\bk=0$ and quadratic at small momenta, i.e. $\omega_\bk = J_1 a_m^2 k^2$ for $\bk \rightarrow 0$. 
The coupling of these spin waves to the exciton via the self-energy $\Sigma_2$
can now be calculated in exactly the same way as for the AFM described in Sec.~\ref{AFM}, but now with the coherence factors $u_\bk = 1$ and $v_\bk = 0$. 
We shall  demonstrate below that it follows from the low energy linear and quadratic spin wave spectra of the AFM and FM
 that their effects on the exciton spectrum exhibit qualitative differences.

\subsection{Self-consistent Born approximation}\label{SCBA_sec}
As we saw above, the motion of the exciton leads to the emission and annihilation of  spin waves in the in-plane magnetic state of the \moire holes. This is closely analogous  to the motion of a hole in an AFM background at half filling, which has been  studied intensely for decades and is 
 relevant to high temperature and unconventional superconductors~\cite{Keimer2015,Wen2011}. 
An important  result of this large body of research is that a single hole moving in an AFM background is 
 accurately described within the $t$-$J$ model by the so-called self-consistent Born approximation 
(SCBA)~\cite{Kane1989,Martinez1991,Liu1991,Diamantis_2021,Nielsen2021}. Remarkably, this holds even for strong interactions, triangular lattices~\cite{Kraats2022}, and non-equilibrium  dynamics~\cite{Nielsen2022}. 

\begin{figure}[ht!]
\centering
\includegraphics[width=1\columnwidth]{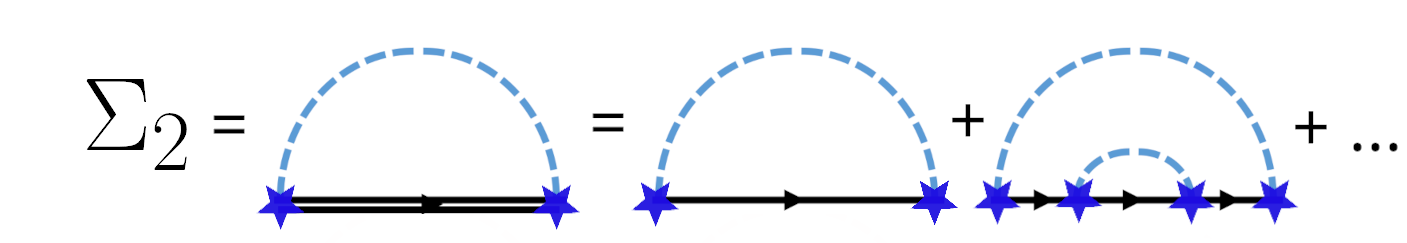}%
\caption{The self-consistent Born approximation for $\Sigma_2$. The double line is the full exciton Green's function. 
}
\label{FigSCBA}
\end{figure}

Due to the close connections  with  the motion of  a hole in an AFM, we expect that an exciton interacting with an in-plane magnetically ordered state 
 is also accurately described by the  SCBA. We therefore adopt this approach to the problem at hand, which  amounts to using the self-consistent exciton Green's function 
\begin{equation}
{\mathcal G}({\mathbf p}',{\mathbf p},i\omega_n)=\frac1{i\omega_n-\epsilon^x_{\bp}-\Sigma({\mathbf p}',{\mathbf p},i\omega_n)}
\label{Greensfn}
\end{equation}
when evaluating Eq.~\eqref{Selfenergy2} with $\Sigma({\mathbf p}',{\mathbf p},i\omega_n)=\Sigma_1({\mathbf p}',{\mathbf p})+\Sigma_2({\mathbf p}',{\mathbf p},i\omega_n)$. This approach is illustrated diagrammatically in Fig.~\ref{FigSCBA}. The resulting self-energy term $\Sigma_2$ given by Eq.~\eqref{Selfenergy2} 
has the same mathematical structure as in case of a single hole moving in an AFM background described by the SCBA~\cite{Kane1989,Martinez1991,Liu1991,Diamantis_2021,Nielsen2021}.
Note that the SCBA goes beyond the second order by summing a subset of terms to infinite order in the exciton-hole scattering events, characterised by the so-called rainbow diagrams shown in Fig.~\ref{FigSCBA}. In this way, SCBA is able to quantitatively account for  
strong interaction effects regarding hole motion in AFMs, and we expect the same to be the case for the motion of an exciton strongly interacting with in-plane magnetically ordered \moire system. 

\section{Numerical results} \label{NumSec}
Having set up a strong coupling formalism for an exciton interacting with the \moire holes forming  in-plane magnetic order, we are now ready to  present  results obtained by numerically solving Eqs.~\eqref{Selfenergy1}, \eqref{Selfenergy2}, and  \eqref{Greensfn} self-consistently. The spin-spin correlation function in Eq.~\eqref{Selfenergy2} is calculated from Eq.~\eqref{SzSz} using 
the spin wave spectrum and coherence factors for either the AFM or the FM. 
Motivated by recent experiments~\cite{Tan2020}, we take $\epsilon_T = -25$ meV
and $\Delta = 8$ meV so that the pole of the ${\mathcal T}$-matrix is at 
$\epsilon_p(\bk) \sim -17$ meV for small momenta. Since the 
spin wave energies are of the order $J_1\sim 0.1$ meV and, as shown below, the relevant exciton energies are in the range of a few meV, we can safely neglect the frequency dependence of the exction-hole scattering matrix, which justifies the approximation introduced 
 in Sec.~\ref{ScatteringSec}. 
Finally,  we use $t \sim 3$ meV for the hole tunneling parameter between the \moire and exciton layers, which is an experimentally 
realistic value as discussed in App.~\ref{tunn_sec}, and consider zero-temperature. Details of the numerical evaluation of the exciton self-energy using the SCBA are provided App.~\ref{numerics_sec}.

\subsection{Anti-ferromagnet}
\begin{figure}[ht!]
\centering
\includegraphics[width=1\columnwidth]{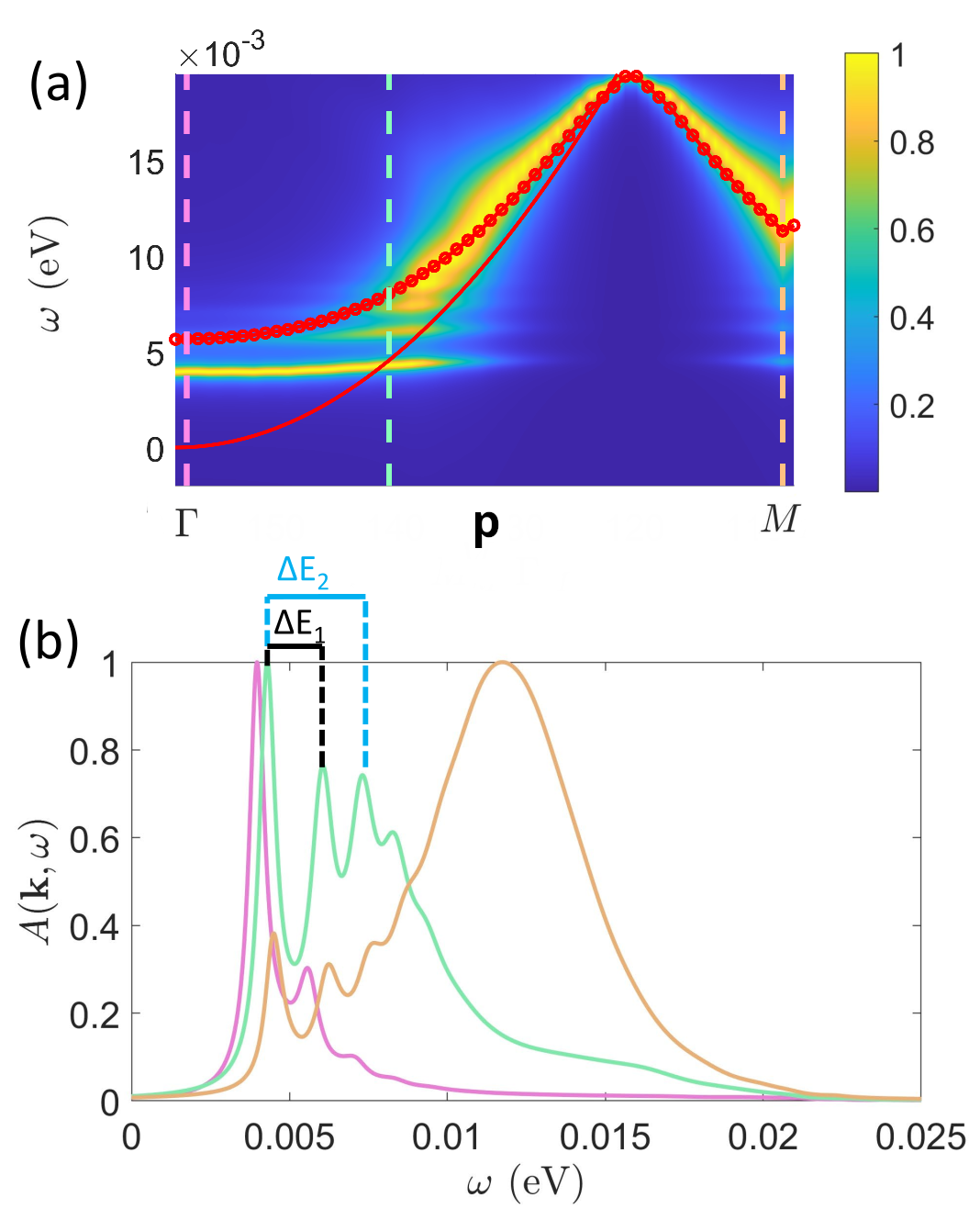}
\caption{Exciton spectral function $A(\bp,\omega)$ for the twist angle  $\theta=3^\circ$  and an in-plane AFM.
 Panel (a) presents $A(\bp,\omega)$ as a function of the frequency and  momentum along the straight path connecting the $\Gamma$ and $M$ points in the mBZ.
 The red solid/dotted line gives the exciton dispersion in the absence/presence of the Umklapp scattering term $\Sigma_1$ but without the $\Sigma_2$ term. Panel (b) shows the spectral function as a function of  frequency for fixed momenta given by the vertical lines  in (a).}
\label{FigRes1}
\end{figure}

We first discuss the case of a \moire AFM by using the corresponding spin wave spectrum and coherence factors with a  
value of $J_1$ obtained as in Sec.~\ref{Moiresystem}. 
Figure \ref{FigRes1}(a) shows the diagonal exciton spectral function $A(\bp,\omega) = -2\text{ Im}[\mathcal{G}(\bp,\bp,\omega +i0^+)]$  obtained from the SCBA 
 as a function of energy $\omega$ and momentum $\bp$ for the twist angle $\theta=3.0\degree$. 
For comparison, we also plot in Fig.~\ref{FigRes1}(a) the  non-interacting exciton 
dispersion $\epsilon^x_{\bk}=k^2/2m_x$  as a solid red line and the dispersion obtained including only the Umklapp term $\Sigma_1$ in the exciton self-energy as a red dotted line. 
This shows that the Umklapp scattering increases the exciton energy for small momenta with respect to its non-interacting energy. It can be understood from the fact that the exciton-hole scattering matrix effectively corresponds to a repulsive interaction, since the trion energy  $\epsilon_T = -25$ meV is well below any relevant energy. 
As detailed in App.~\ref{AppsecSigma1}, one can show that the zero-momentum energy shift due to the Umklapp potential is of the order  $\mathcal{O}(t^2Z/\Delta^2 |\epsilon_T| a_m^2)$.

Consider next the spectral function when the full self-energy $\Sigma_1+\Sigma_2$ is used, which takes into account the scattering on the spin waves 
to infinite order through the self-energy  term $\Sigma_2$. First, we see from 
Fig.~\ref{FigRes1}  that the spectral function exhibits a sharp peak for small momenta. This peak corresponds to a 
 quasi-particle consisting of the exciton dressed by  spin waves in the \moire AFM as illustrated in Fig.~\ref{Fig1}(a), and it quantitatively confirms the prediction regarding its existence in Sec.~\ref{phys_sec}. We denote this quasipaticle as an exciton-polaron in 
analogy with   what has been done for an exciton coupled to electrons or other distinguishable excitons~\cite{Sidler2017,tan2022bose}. In contrast to the Umklapp term $\Sigma_1$, $\Sigma_2$ decreases  the energy $\epsilon_{\bp}^P$ of the exciton-polaron.
This follows from the fact that  $\Sigma_2$ describes the coupling to higher energy states 
containing  spin waves as described through the SCBA. 
 We moreover see that the dressing by the spin waves significantly increases the effective mass $m^*$ of the quasi-particle making its dispersion flat. 
 
 For larger momenta, Fig.~\ref{FigRes1}(a) shows that the quasiparticle peak becomes significantly broadened corresponding to a 
  short lived quasiparticle. Physically, this decay sets in when the energy of the exciton-polaron $\epsilon_{\bp}^P$ is sufficient to scatter resonantly on the spin wave spectrum, i.e.\  when $\epsilon_{\bp}^P \ge  \min_\bq [ \omega_\bq + \epsilon_{\bp-\bq}^P]$.  
Figure \ref{FigRes1}(a) furthermore shows that when the damping sets in, the maximum of the broad continuum approaches the  Umklapp-dispersion of the exciton.
 The presence of the sharp quasi-particle peak for small momenta and a broadened spectrum for larger momenta is further illustrated 
in Fig.~\ref{FigRes1}(b), which plots the  spectral function as a function of $\omega$ for a number of selected momenta  indicated by the vertical dashed lines in Fig.~\ref{FigRes1}(a).

\begin{figure}[ht!]
\centering
\includegraphics[width=1\columnwidth]{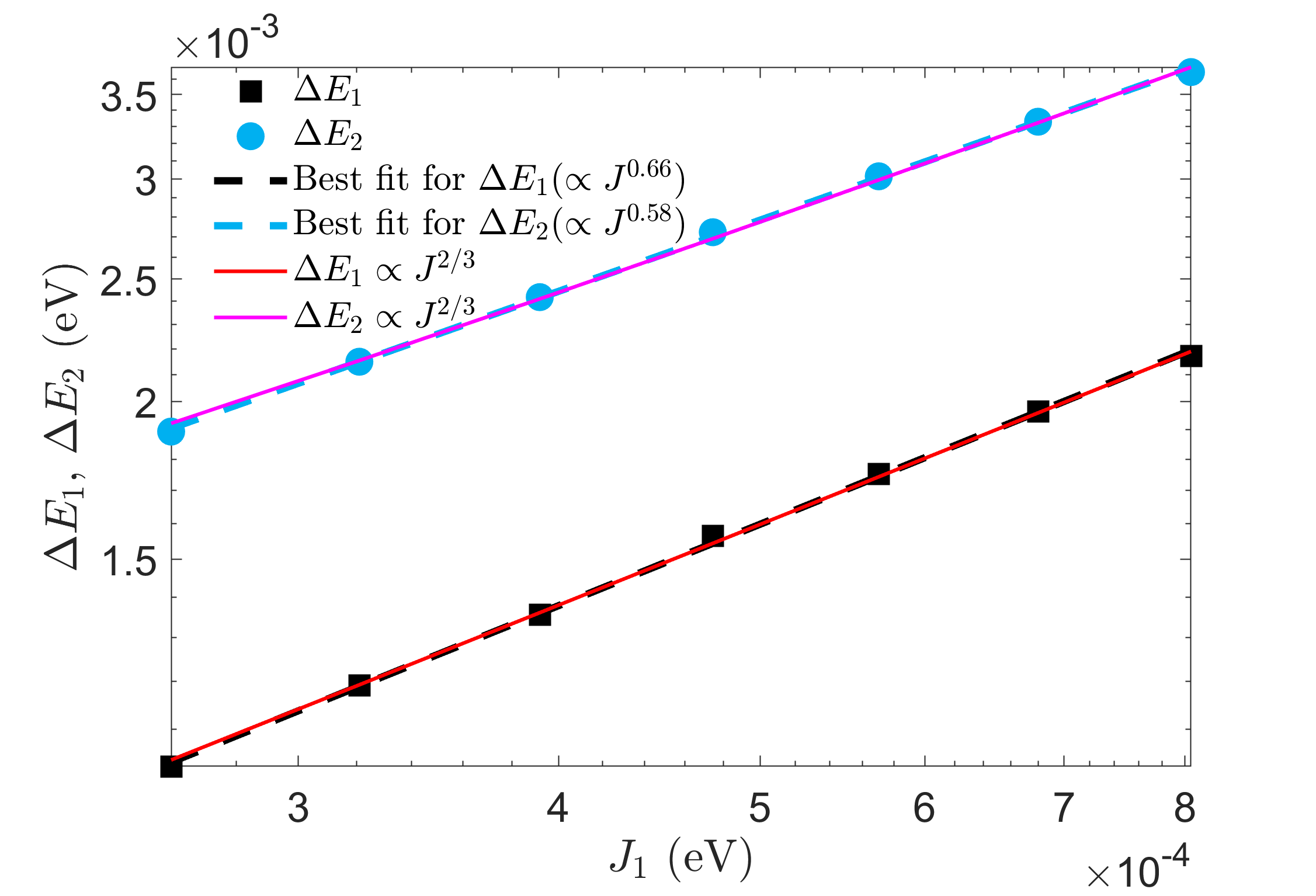}
\caption{Log-log plot of the two lowest excitation energies  $\Delta E_1$ and $\Delta E_2$ (see Fig.~\ref{FigRes1}) as a function of $J_1$ for the 
momentum $\bp=0.35 \bk_M$. Black squares/blue circles are numerical results, and the dashed lines show the power law fits 
 $\Delta E_i = \beta_i J^{\alpha_i}$ where $\alpha_i$ and $\beta_i$ are the fitting parameters. For $\Delta E_1$ and $\Delta E_2$, we obtain $\alpha_1 = 0.66$ and $\alpha_2=0.58$. The solid lines show the geometric string prediction  $\Delta E_i = \gamma_i J^{2/3}$ with $\gamma_i$ a fit parameter.  The variation in the 
 nearest neighbour spin coupling $J_1$ corresponds to sweeping the twist angle from $2.6\degree$ to $3.2\degree$.}
\label{FigRes2}
\end{figure}

In addition to the sharp quasi-particle  peak, Fig.~\ref{FigRes1} shows
a number of intriguing broader peaks appearing at higher energy for small momenta. They 
 correspond to excited states with an energy and decay rate  given by the peak center and width respectively. 
To analyse their physical origin, we plot 
 in Fig.~\ref{FigRes2} the two lowest excitation energies $\Delta E_1$ and $\Delta E_2$ (see Fig.~\ref{FigRes1}) as a function of $J_1$ 
 for the momentum $\bp=0.35 \bk_M$. In addition to the numerical data, we present as the dashed lines the best power-law fits $\Delta E_i \propto J_1^{\alpha_i}$, where $\alpha_1 = 0.66$ and $\alpha_2 = 0.58$ for $\Delta E_1$ and $\Delta E_2$, respectively. 
 These fits are  close to a  $J_1^{2/3}$ scaling expected for the so-called geometric string excitations~\cite{Liu1992,Trumper2004,Hamad2008,Manousakis2007},
 which is therefore also plotted  in Fig.~\ref{FigRes2}, showing a very good agreement   with the numerical results.

By following the the same logic   as for the case of a hole hopping in an  AFM background~\cite{Liu1992,Trumper2004,Hamad2008,Manousakis2007}, such a scaling of  $\Delta E_1$ and $\Delta E_2$  can be taken as a fingerprint of geometric string excitations.   
As we discussed in Secs.~\ref{phys_sec} and \ref{AFM} and illustrated in Fig.~\ref{Fig1}(b), the exciton  leaves a trail of spin flips in the \moire lattice sites as it 
 moves around. 
 Since these spin flips cause an energy penalty of the order of $J_1$, the exciton experiences an effective linear potential giving 
rise to Airy-like eigenstates with the corresponding eigen-energies proportional to $ J_1^{2/3}$. 
A very attractive feature of the present setup is that one can tune $J_1$ by changing the twisting angle of the \moire bi-layer. This 
allows one to experimentally verify the smoking gun $J_1^{2/3}$ energy scaling of string excitations, 
which has turned out to be very difficult to realize for holes in  AFMs. In fact, the range of 
$J_1$ values shown in Fig.~\ref{FigRes2} corresponds to scanning the twist angle 
from $\theta = 2.6^\circ$ to $\theta = 3.2^\circ$ in the MoSe$_2$-WS$_2$ bilayer. 
Hence, our results show how the flexibility of TMDs may allow for a detection of the elusive string excitations. 
Finally, we emphasize that it is 
essential to use the non-perturbative SCBA to describe the string excitations. Indeed, as is shown explicitly  App.~\ref{2nd_ord_app}
 they are completely missed by   second order pertubation theory.

In Fig.~\ref{FigGKKG} in App.~\ref{numerics_sec}, we plot the exciton spectral function along other high symmetry directions in the mBZ. It exhibits the same features, 
i.e.\ a clear exciton-polaron quasiparticle peak and broader peaks coming from excited geometric string states, which demonstrates that these states  appear throughout the mBZ. 

\subsection{Ferromagnet}
\begin{figure}[ht!]
\centering
\includegraphics[width=0.9\columnwidth]{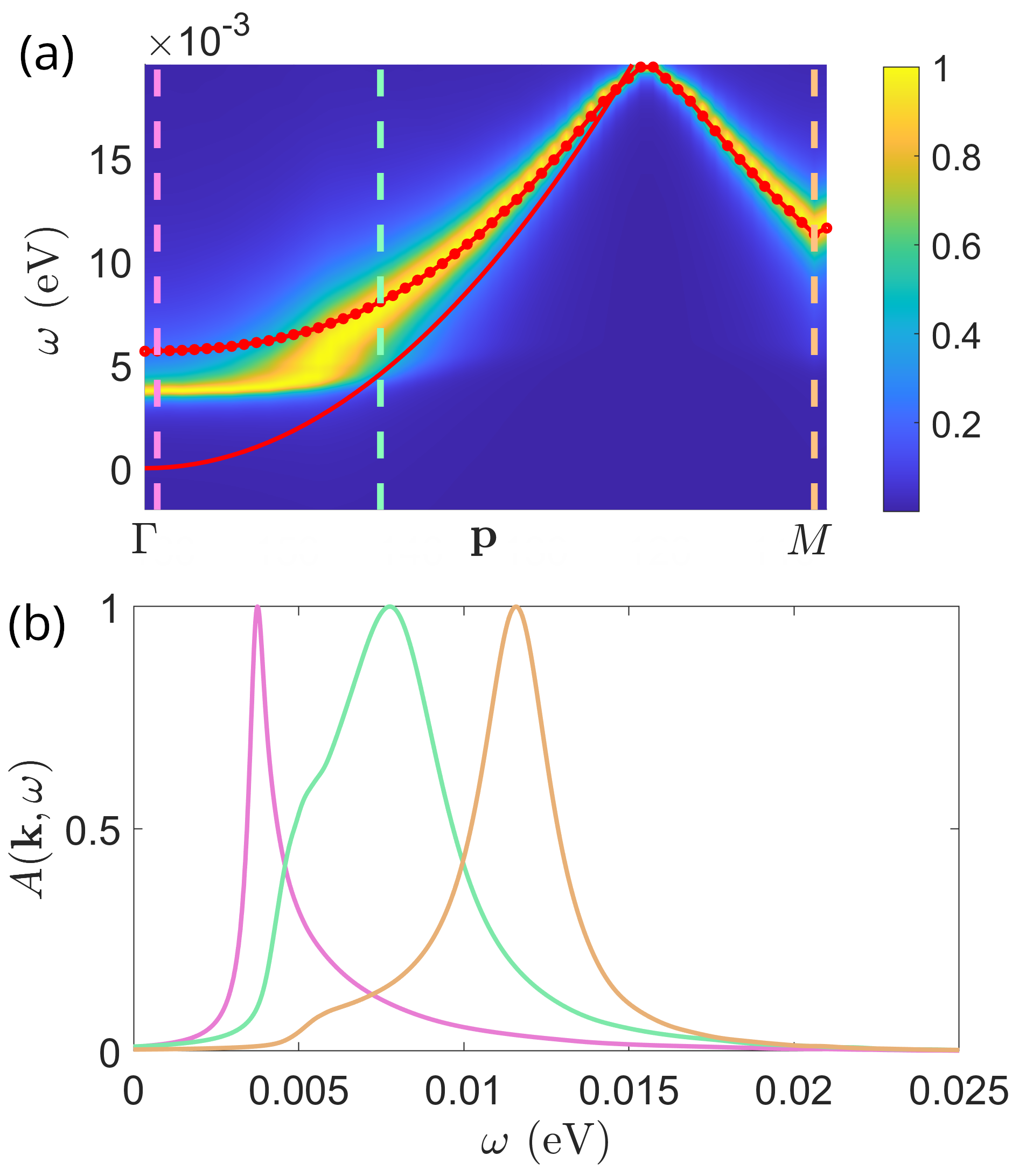}
\caption{Exciton spectral function $A(\bp,\omega)$ for $\theta=3^\circ$  for an in-plane FM.
 Panel (a) presents $A(\bp,\omega)$ as a function of the frequency and  momentum along the path connecting the $\Gamma$ and $M$ points in the mBZ.
 The red solid/dotted line gives the exciton dispersion in the absence/presence of the Umklapp scattering term $\Sigma_1$ but without the $\Sigma_2$ term. Panel (b) shows the spectral function as a function of  frequency for fixed momenta given by the vertical lines  in (a).}
\label{FigRes4}
\end{figure}

We now turn to the case where the \moire holes form an in-plane FM, which we describe   by  flipping the sign of $J_1$. In Fig.~\ref{FigRes4} we show the resulting exciton spectral 
function for $J_1= -0.54$ meV. As for the case of an AFM, we see a clear exciton-polaron quasiparticle with energy above the bare exciton but below that predicted by the Umklapp scattering, and with a large effective mass. 
This quasiparticle furthermore becomes strongly damped for larger momenta due to scattering on the spin waves, where its dispersion approaches the exciton energy given by  the
Umklapp scattering. 

Compared to the case of an AFM, the most striking difference is the absence of  string excitations in the exciton spectrum. This is somewhat
surprising since the basic ingredients leading to the presence of such excitations seem to be present for the FM too, i.e.\  
the exciton leaves a string of magnetic frustration in its wake leading to a linear potential. We conjecture  that this string picture is
nevertheless invalid for the FM, since an 
initially localised magnetic frustration rapidly spreads out in  real space due to the quadratic low energy dispersion of the spin waves. This aspect is discussed 
further  in App.~\ref{string_ex_sec}.

\section{Discussion and outlook} \label{ConcSec}
In this work, we developed a general theory for the coupling between an exciton in a TMD monolayer 
and the spin and charge degrees of freedom in an adjacent \moire lattice formed by a TMD bilayer. Using this theory combined with a non-perturbative  self-consistent Born approximation, we  explored the  coupling between the exciton 
and spin waves in a half-filled \moire lattice hosting an in-plane AFM or FM order. We showed that this gives rise to the 
formation of quasiparticle states (polarons) consisting of the exciton surrounded by magnetic frustration in the \moire lattice, damping, as well as geometric string states.

The predicted emergence of a well-defined polaron adds to the list of interesting quasiparticles that can be realised with excitons in TMDs, which includes the 
 experimentally observed Fermi and Bose polarons~\cite{Sidler2017,tan2022bose}. The polarons in the present paper are closely related to the magnetic 
polarons consisting of a mobile hole in a AFM lattice, which play a fundamental role for charge transport in 2D quantum materials including 
high temperature and unconventional superconductors~\cite{Keimer2015,Wen2011}. We note that as opposed to these materials, the flexibility of TMD bi-layers 
allows one to tune the nearest neightbor spin-spin coupling constant by changing the twist angle. This opens up the possibility to observe 
the elusive string excitations via their the smoking gun energy scaling.

Since we predict clear and observable changes of the exciton spectrum, our results 
have intriguing perspectives for using excitons as much needed quantum sensors  for the \emph{spin} correlations of 
the rapidly growing class of  2D van der Waals materials. It has previously been shown how the first order mean-field  term $\Sigma_1$ gives rise
to  Umklapp scattering branches in the exciton spectrum signalling Wigner crystalisation~\cite{Smolenski2021} as well as 
out-of-plane magnetic order~\cite{Salvador2022}. Here, we demonstrate that higher order terms 
included in $\Sigma_2$ couple the exciton to spin correlations, which can be in any direction and do not have to exhibit long range order. Also, it was   
recently shown that the in-plane AFM order increases the effective mass of an exciton  residing in the \moire 
lattice~\cite{Huang2022}, and that the spatial periodicity related to different charge and magnetically ordered states  
influences the plasmon spectrum of \moire systems through the band folding~\cite{Papaj2022}. Finally, the possible magnetic ground states of a
RuCl$_3$ monolayer were shown to change due to the hybridisation with an adjacent graphene layer and giving rise to graphene magneto-resistance~\cite{Shi2023}. 
Major strengths of our probe setup include that it gives  access to spin correlations, which is essential for  identifying 
interesting phases with no long-range order, and that the exciton probe is non-evasive. 

Experimentally, the excitons close to the $\Gamma$ point can be probed by optical means where a resolution 
of the order of a few meVs has been achieved in TMDs~\cite{Smolenski2021,Salvador2022}. This is sufficient to clearly 
observe the spectral features we predict without significant  broadening, which could otherwise be significant due to a finite exciton lifetime for nearly degenerate conduction bands~\cite{Marinov2017}. 
The spectral features away from the $\Gamma$ point are
 accessible using electron energy-loss spectroscopy (EELS), which  has recently been used to measure the exciton spectrum in a mono-layer WSe$_2$~\cite{Hong2020}. 
 While the energy resolution achieved in this  experiment for non-zero momenta is not sufficient to detect the spectral features we predict 
 such as the string states, the experimental techniques are improving rapidly and much higher precision is indeed 
 expected in the future~\cite{Hong2020}. 
 We also note that since the effects on the exciton spectrum increase with the exciton-electron(hole) interaction 
 strength, they can be much larger and therefore more easily resolved experimentally than shown in Figs.~\ref{FigRes1} and \ref{FigRes4}. 
Finally, the spectral signal from the probe and \moire layers can easily be distinguished as they are well separated in energy~\cite{Xu2020}.

One  interesting phase with  no long range order but subtle and non-trivial correlations is the Dirac spin liquid, which is a strong candidate for the ground state of the triangular $J_1-J_2$ Heisenberg model in certain parameter regimes~\cite{Sherman2023,Drescher2022}. It would be interesting to explore how  the exciton  couples to the low energy
degrees of freedom of this phase such as spinon 
excitations, and in what ways this shows up in the exciton spectrum.  
 Spin liquids should be within experimental reach  in TMD bi-layers since they naturally realise a triangular lattice, and since the ratio $J_2/J_1$ is tunable via the twist angle and the screening of the Coulomb interactions~\cite{Morales-Duran2022}. There likely are many other interesting effects to observe such as for instance a cascade of phase transitions as seen in magic angle graphene~\cite{Zondiner2020}.

The results presented here open up several other promising research directions. One can use the Feshbach resonance mediated by the trion state to increase the interaction 
between the exciton and the \moire holes and thereby the effects on the exciton spectrum. This is achieved by tuning the trion energy via the energy  offset $\Delta$~\cite{Schwartz2021,Kuhlenkamp2022}. Since the static approximation for the exciton-hole  scattering matrix used in the present 
paper is insufficient to analyse this, one must instead keep the full frequency dependence, which is  a 
 a technically  challenging problem. Our results also 
raise the  fundamental questions regarding what properties of the magnetic environment are necessary to stabilise geometric string exctitations.  
Another interesting research topic is how to use excitons pinned in a \moire lattice, as has been achieved experimentally~\cite{Wu2017,Yu2017,Zhang2021}, 
to create probes with spatial resolution. Pinning several excitons could furthermore enable powerful and highly useful multiplex sensors 
capable of simultanuously measuring electron/hole correlations at two or more spatial positions. The hybridisation of the excitons with photons in an optical cavity to 
form polaritons~\cite{Carusotto2004} would have several interesting several effects such as the quantum state of the \moire system being imprinted on the outgoing light. 
Finally,   we considered in the present paper a single exciton, which does not change the quantum state of the \moire lattice in the 
thermodynamic limit. Experimentally, this corresponds to the limit where the density 
of excitons is much smaller than the electron density in the \moire lattice.
This basic idea has also been used by many experimental groups exploring a few mobile 
 impurity atoms forming so-called polarons by scattering on atoms in their  neighborhood in a large surrounding quantum degenerate atomic gas~\cite{Massignan2014}. 
 It would be interesting in the future to consider different exciton and electron concentrations, which will realise new and intriguing Bose-Fermi mixtures.

\begin{acknowledgments} 
This work was supported by the Danish National Research Foundation through the Center of Excellence “CCQ” (Grant no. DNRF156).
\end{acknowledgments}

\appendix

\section{Hole tunneling between the \moire system and the exciton layer}\label{tunn_sec}
 To decrease the hole tunneling between the \moire system and the  TMD monolayer and prevent the formation of a large three-layer moir\'e lattice, we propose to apply a $30^\circ-$rotated  WS$_2$ monolayer as a barrier between the \moire system and the exciton layer. Due to a large rotation angle of the barrier WS$_2$ layer, its energy band structure is highly detuned by an energy off-set $\tilde \Delta $ with respect to the relevant energies of the \moire system and the monolayer MoSe$_2$.  As a result, one can use the Schrieffer-Wolff transformation to estimate the hole tunneling strength between the moir\'e system and the ML MoSe$_2$  as $t\sim \tilde{t}^2/\tilde \Delta$, where $\tilde{t}$ is the interlayer tunneling parameter between the spacer layer (WS$_2$)
 and its two neighbouring MoSe$_2$ layers  [see Fig.~\ref{Fig1}(b)]. The energy off-set of the valence band edges between un-rotated  MoSe$_2$ and WS$_2$ layers is around $V\sim 270$ meV~\cite{Ruiz-Tijerina2019}. Moreover, the rotation by $30\degree$ shifts the energies of the electron states of WS$_2$ that momentum-match with the relevant states of the neighbouring MoSe$_2$ layers, by around $\sim 1$ eV~\cite{Terrones2013}. Hence, we get a rough estimate $\tilde \Delta \sim 0.7$-$0.8$ eV. On the other hand, if we use $\tilde t \sim 50$ meV given in Ref.~\cite{Ruiz-Tijerina2019}, we have $t = \frac{\tilde{t}}{\tilde \Delta} \sim 3$ meV. This value can be  tuned significantly by for example choosing some other material for the spacer layer or by changing the rotation angle of the spacer.

\section{\Moire Hamiltonian}\label{moiresec}
In this section, the \moire Hamiltonian for electrons is derived in case of a hetero-bilayer by using the continuum model deployed in Refs.~\cite{Ruiz-Tijerina2019,Alexeev2019}. In this model, the long-wavelength \moire potential arises due to  interlayer tunneling processes.

\begin{figure}
  \centering
    \includegraphics[width=0.7\columnwidth]{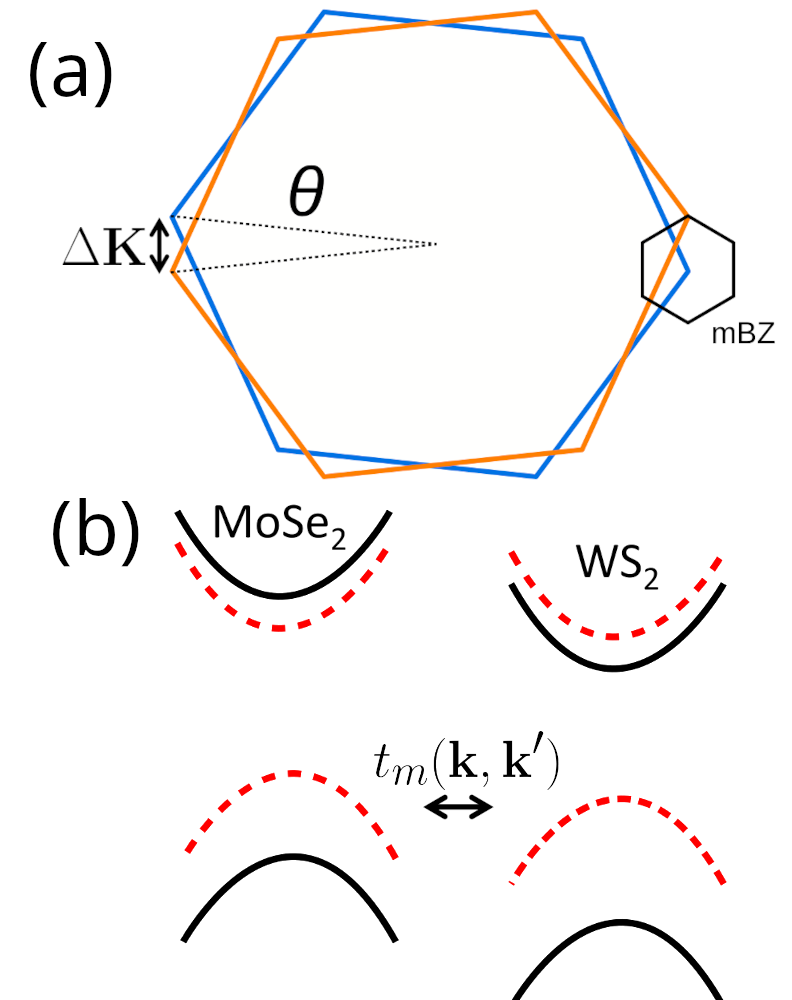}
    \caption{ (a) Schematic of the original BZ of the two layers forming the \moire lattice (blue and orange hexagons) as well as the \moire BZ (mBZ) of the $K$-valley. (b) Schematic picture of the highest valence and lowest conductance bands of the monolayer MoSe$_2$ and WS$_2$ at the $K$-valley.  Dashed red (solid black) lines correspond to spin-down (spin-up) bands. The \moire potential arises due to a finite interlayer tunneling $t_m(\bk,\bk')$. In our analysis we only consider the highest valence band degrees of freedom. }
   \label{Fig_S1}
\end{figure}

We consider a TMD heterobilayer  and label the two layers as L1 and L2. For concreteness, we take L1 to be a MoSe$_2$ and L2 a WS$_2$ monolayer.  The MoSe$_2$ and WS$_2$  monolayers have a hexagonal Brilliouin zone (BZ) and feature direct band gaps at the $K$ and $K'$-valleys. For two layers, the $K$-valleys are at $\bk = [4\pi/(3 a_1),0] \equiv \textbf{K}_1$ and $\bk = [4\pi/(3 a_2),0] \equiv \textbf{K}_2$ with $a_i$ being the original lattice constants. The $K$-points of the two layers differ from each other by $\Delta \bkk = \bkk_2 - \bkk_1$. 
A finite lattice mismatch ($a_1 \neq a_2$) or a possible relative twist angle $\theta$ between the layers yields $\Delta \bkk \neq 0$. This is illustrated in Fig.~\ref{Fig_S1}(a) in case of a finite twist angle. When the lattice mismatch and the twist angle is small, we have $|\Delta \bkk| << |\bkk_1|$. In this case, the interlayer tunneling leads to the hybridization between the low-energy electronic states of the two layers, separately in the $K$ and $K'$-valleys. Correspondingly, the system acquires long-wavelength moire pattern described by the \moire periodicity $a_m$ and the 
reduced \moire Brilliouin Zone (mBZ), see Fig.~\ref{Fig_S1}(a). Due to the large momentum mismatch between the original $K$ and $K'$-valleys, the $K$-valley states do not hybridize with the $K'$-valley degrees of freedom and hence, as the TMD monolayers feature spin-valley locking due to the intrinsic spin-orbit coupling~\cite{Wang2018,Mueller2018}, the valley index corresponds to the $z$-component of the spin. We can therefore consider the one-particle \moire Hamiltonian for each valley separately. We denote the valley-index as $\sigma \in \{\uparrow,\downarrow \}$, with $\sigma = \uparrow$ ( $\sigma = \downarrow$) referring to the $K$-valley ($K'$-valley) and express the corresponding one-particle Hamiltonian as  $H^\sigma_0$. 
We consider here only the $K$-valley electrons; the corresponding result for the $K'$-valley is obtained from the $K$-valley via  time reversal. 

The valence and conduction band dispersions of the decoupled  MoSe$_2$ and WS$_2$ monolayers  at the $K$-valley are shown schematically in Fig.~\ref{Fig_S1}(b). To a good approximation, these band structures can be taken to be parabolic. Furthermore, the valence and the conduction bands are separated by a large band gap. As it is common in experiments, we take the \moire system to be hole-doped within the highest valence band and hence we discard all the other bands. The highest valence bands of the two layers are coupled and the \moire pattern is formed due to the interlayer electron tunneling term $t_m(\bk,\bk') = t_m [\delta_{\bk,\bk'} + \delta_{\bk - \bk', \bb^m_{1}} + \delta_{\bk - \bk', \bb^m_{2}}]$ which describes a tunneling process from momentum state $\bk'$ to $\bk$ with $\bb^m_i$ being the basis vectors of the \moire reciprocal lattice~\cite{Ruiz-Tijerina2019,Wang2017}. 
Consequently, the continuum Hamiltonian for the bi-layer  reads
\begin{align}
H^\uparrow_0 &= \sum_{\bk \in \textrm{mBZ}} \Psi^\dag(\bk) \mathcal{H}(\bk) \Psi(\bk)\nonumber \\
  \Psi(\bk) &= [c_{1}(\bk), c_{2}(\bk - \Delta\bkk)]^T \nonumber \\
 [c_{i}(\bk)]_\alpha &= c_{i\uparrow}(\bk + \bgg_\alpha) \nonumber \\
\mathcal{H}^\uparrow(\bk) &= \begin{bmatrix}
	\mathcal{H}^\uparrow_{1}(\bk) &   \mathcal{H}^\uparrow_{12}(\bk) \\
	 [\mathcal{H}^\uparrow_{12}(\bk)]^\dag &  \mathcal{H}^\uparrow_{2}(\bk - \Delta\bkk) \label{mHam} 
\end{bmatrix}
\end{align}
Here $c_{i\uparrow}(\bk)$ annihilates an electron within the highest valence band in valley $\sigma=\uparrow$ and layer $i$ ($i=1$ for MoSe$_2$ and $i=2$ for WS$_2$). The reciprocal lattice vectors are $\bgg_\alpha \equiv n_\alpha \bb^m_1 + m_\alpha \bb^m_2$ with $n_\alpha$ and $m_\alpha$ being integers. Moreover, $[\mathcal{H}^\uparrow_i(\bk)]_{\alpha\beta} = \delta_{i,1}V -\delta_{\alpha\beta} \hbar^2(\bk + \bgg_\alpha)^2/2m_{i}$ contains the original parabolic dispersions for layer $i$ with the effective mass $m_i$ and the valence band edge off-set $V$ between two layers. Finally, the off-diagonal block $\mathcal{H}^\uparrow_{12}$ describes the interlayer tunneling $t_m$ between relevant momenta. For $m_i$ and $V$ we adopt the numerical values used in Ref.~\cite{Ruiz-Tijerina2019}, i.e. $m_1 =0.44m_e$, $m_2 =0.32m_e$, $V = 270$ meV where $m_e$ is the bare electron mass. The value used for $t_m$ is discussed at the end of this section.

By diagonalizing the Hamiltonian~\eqref{mHam}, one obtains the \moire band dispersion $\epsilon^\uparrow_{n\bk}$ such that 
\begin{align}
H^\uparrow_0 = \sum_{\bk \in \text{mBZ},n} \epsilon^\uparrow_{n\bk}  \gamma^\dag_{n\bk \uparrow}\gamma_{n\bk \uparrow}    
\end{align}
Here $n = 1,2,...$ is the \moire band index with $n=1$ referring to the highest \moire valence band and $\gamma_{n\bk\uparrow}$ are the corresponding \moire annihilation operators. In Fig.~\ref{Fig_disp} the \moire dispersion $\epsilon^\uparrow_{n\bk}$ is shown for $\theta=2.5^\circ$ computed with our parameters. The \moire hole band operators in Eq.~\eqref{projection_main} then read $h_{n\bk\sigma}^\dag = \gamma_{n -\bk\sigma}$. 

The diagonalization of $H^\uparrow_0$ yields the  following transformation between the Bloch band basis and the original electron operators
\begin{align}
\Psi_\uparrow(\bk)=
\begin{bmatrix}
    U_{\bk \uparrow} \\ V_{\bk \uparrow}
\end{bmatrix}
\gamma_\uparrow(\bk) \equiv B_\uparrow (\bk) \gamma_\uparrow(\bk),
\end{align}
where $[\gamma_\uparrow(\bk)]_n = \gamma_{n\bk\uparrow}$. Furthermore, the \moire Bloch states are stored as the column vectors of the unitary matrix $B_\uparrow (\bk)$ and its components $U_{\bk\uparrow}$ and $V_{\bk\uparrow}$ describe the projections to the two layers. Namely, the original electronic operators in the MoSe$_2$ layer can be expressed as $c_{1\uparrow}(\bk + \bgg_\alpha) = \sum_{n=1} [U_{\bk \uparrow}]_{\alpha n} \gamma_{n\bk\uparrow}$. In other words, the Bloch functions in Eq.~\eqref{projection_main} are  defined as $u_{n\sigma}(\bk+\bgg_\alpha) \equiv [U_{\bk \sigma}]_{\alpha n}$. Likewise, $V_{\bk\uparrow}$ yields the projection to the $WS_2$ layer, i.e. $c_{2\uparrow}(\bk + \bgg_\alpha) = \sum_{n=1} [V_{\bk \uparrow}]_{\alpha n} \gamma_{n\bk\uparrow}$. 


As discussed in the main text, the highest \moire valence band can be described by the effective tight-binding Hamiltonian with appropriate Coulomb interaction terms, as written  in Eq.~\eqref{Hm}. The hopping terms $t_{ij}^\sigma$ are easy to obtain from the dispersion relation of the \moire band, i.e. one has $t_{ij}^\sigma = - \frac{1}{N}\sum_{\bk \in \textrm{mBZ}} \epsilon_{1\bk}^\sigma e^{-i \bk \cdot (\brr_i -\brr_j)}$, where $\brr_i$ is the spatial coordinate of the $i$th \moire lattice site. To obtain the interaction terms, one needs appropriate expressions for the Coulomb interaction $V_{C}(\bq)$. For the interlayer and intralayer  interaction vertices we have in 
momentum space $V_{C,s}(q) =  \frac{e^2}{2 q \epsilon_s(q)} \tanh(q d_g)$, where $s = \perp$ ($s= \parallel$) stands for the interlayer (intralayer) interaction. We have furthermore taken into account the metallic gates, separated by a distance $d_g$ from the \moire sample, which screen the Coulomb interactions via the hyperbolic tangent function~\cite{Chubukov2017}.  The permittivities can be derived from the system geometry and in our case are given by $\frac{1}{\epsilon_{\parallel}(q)} = \frac{1 + r^* q - r^* q e^{-2qd_l}}{\epsilon_0 \epsilon_r [(1 + r^* q)^2 - (r^* q)^2 e^{-2qd_l}]}$ and $\frac{1}{\epsilon_{\perp}(q)} = \frac{e^{-qd_l}}{\epsilon_0 \epsilon_r [(1 + r^* q)^2 - (r^* q)^2 e^{-2qd_l}]}$~\cite{Danovich2018}. Here $\epsilon_r$ is the average dielectric permittivity of the embedding material, $r^*$ is the screening length of a TMD monolayer and $d_l$ is the separation between two TMD layers. In our calculations we have chosen $\epsilon_r = 20$ due to the existence of the spacing and probe TMD layers. Moreover, we have $\epsilon_r r^* \sim 4$ nm~\cite{Ruiz-Tijerina2019},  $d_l = 0.6$ nm and $d_g = 100$ nm.

Once the parameters for Eq.~\eqref{Hm} are obtained, we can compute the parameters of the effective Heisenberg Hamiltonian for the half-filled \moire band by using the well-established strong coupling  theory that maps the half-filled Fermi-Hubbard model to the Heisenberg model~\cite{MacDonald1988}. To this end, we 
employ the expressions given in the Supplementary Material of Ref.~\cite{Morales-Duran2022}, i.e. $J_1 = -4X_{\text{NN}} + \frac{4t_{\text{NN}}^2}{U_0-U_{\text{NN}}}$ and 
$J_2 = -4X_{\text{NNN}} + \frac{4t_{\text{NNN}}^2}{U_0-U_{\text{NNN}}} + \frac{8t_{\text{NN}}^4}{U_0-U_{\text{NN}}^3} (1 - \frac{U_0-U_{\text{NN}}}{U0 - U_{\text{NNN}}} + \frac{U_0-U_{\text{NN}}}{2U_0 - 3U_{\text{NN}} + U_{\text{NNN}}} )$, where subscripts refer to NN or next-nearest-neighbour (NNN) terms. Note that these expressions for $J_1$ and $J_2$ assume real-valued hopping terms. As discussed in the main text, our analysis yields only a small imaginary part for $t_{\text{NN}}$, validating the use of the expressions of Ref.~\cite{Morales-Duran2022}.

Coming back to choosing the interlayer tunneling parameter $t_m$, we note that the value given in Ref.~\cite{Ruiz-Tijerina2019}, i.e. $t_m \sim 50$ meV, is too small in a sense that it yields $J \sim -0.06$ meV for the zero twist angle. However, in experiments~\cite{ciorciaro2023kinetic} it has been shown that for the non-twisted and half-filled MoSe$_2$/WS$_2$, the magnetic order vanishes. This very likely happens due to the fact that the direct ferromagnetic exchange term $X_{NN}$ cancels the antiferromagnetic super-exchange term in the expression of $J_1$. To match this cancellation at $\theta = 0\degree$, we use instead a value of $t_m \sim 90$ meV, which at $\theta = 0\degree$ yields a vanishingly small $J_1$. For larger twist angles used in our calculations, namely $\theta \in [2.6\degree,3.3\degree]$, we find a positive $J_1$, i.e. antiferromagnetic order. 
The fact that we need to fine-tune the parameters of  the continuum model of Ref.~\cite{Ruiz-Tijerina2019} might imply that the predominant origin for the \moire potential of MoSe$_2$/WS$_2$ is not the inter-layer tunneling but the lattice reconstruction which could be addressed by using a continuum model with a one-particle long-wavelength \moire potential instead of a interlayer tunneling of Ref.~\cite{Ruiz-Tijerina2019}. This, however, does not affect the generality of our results as both the continuum models anyway lead to an effective triangular \moire lattice and because we study both  antiferromagnetic and ferromagnetic in-plane orders.

\section{Exciton-hole scattering}\label{Tmat_sec}
In this section, we derive the expression for the scattering matrix beween an exciton in the mono-layer and a hole in the \moire system. To this end, we only consider the scattering between spin $\uparrow$ excitons and spin $\downarrow$ holes and discard the other scattering channel as its effect is  negligible. Correspondingly, in the following discussion we suppress the spin index of the hole for a moment.

\begin{figure}
  \centering
    \includegraphics[width=1\columnwidth]{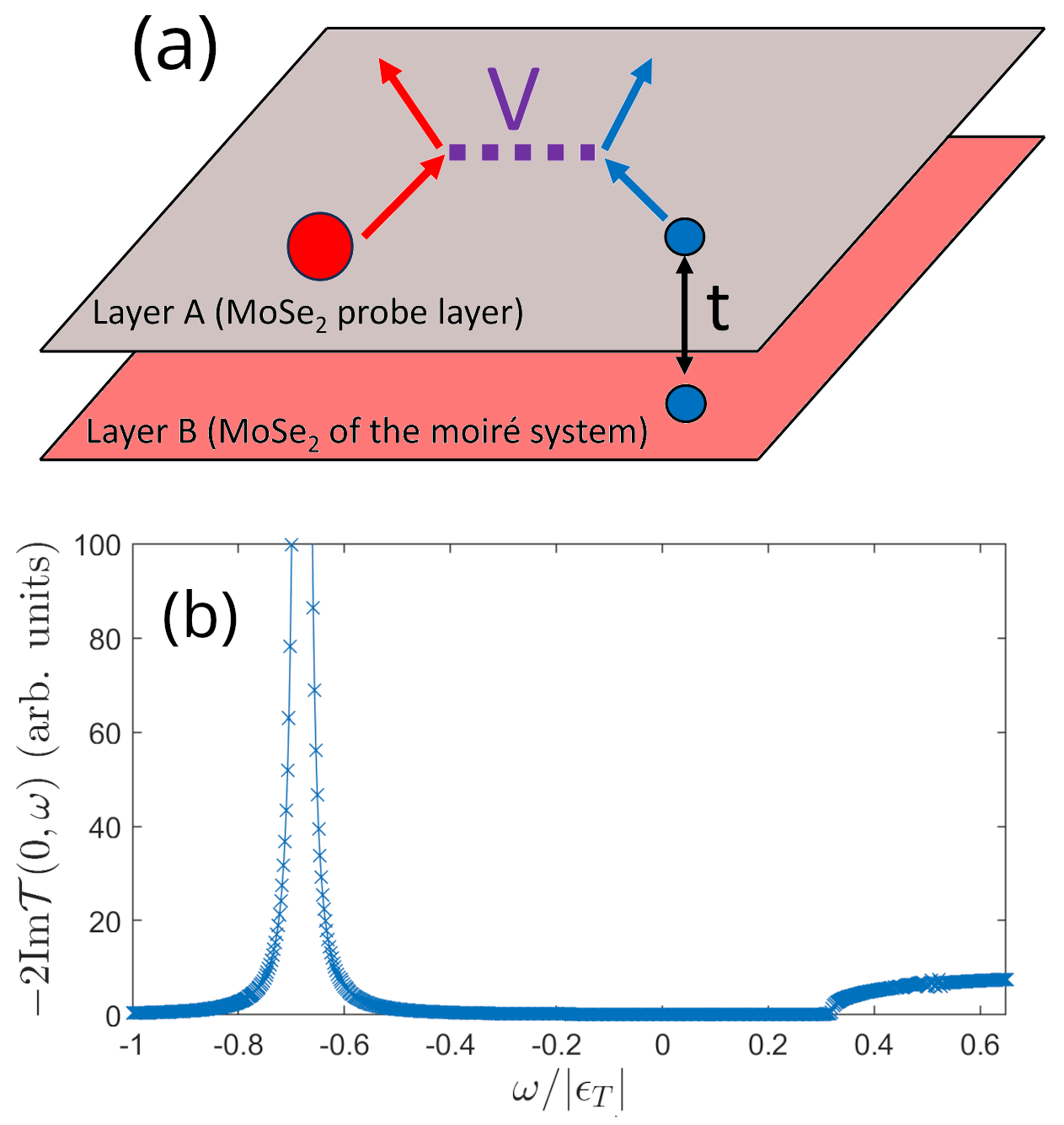}
    \caption{(a) Schematic of the two-layer setup realising  resonantly enhanced scattering between an exciton (red dot) in layer A and a hole (blue dots) in layer B~\cite{Kuhlenkamp2022}. The 
    tunneling between the two layers is $t$ and the interaction between the exciton and the hole is denoted as $V$. (b) Spectral function of the scattering matrix $\mathcal{T}(0,\omega)$ as a function of $\omega$ for $\Delta = \frac{8|\epsilon_T|}{25}$ with finite broadening $\eta = 0.0015 |\epsilon_T|$.}
   \label{two_layer}
\end{figure}

To simplify our analysis, we first consider a bilayer system of two TMD monolayers~\cite{Kuhlenkamp2022}, shown in Fig.~\ref{two_layer}, where the exciton resides in layer A and and a hole can tunnel between the layers A and B with the tunneling strength $t$. Therefore in case of our \moire setup, layer A corresponds to the probe TMD monolayer 
and layer B is the MoSe$_2$ monolayer of the \moire system. The Hamiltonian for the exciton and the hole is written in Eq.~\eqref{FeshbachHam}.
To a good approximation, the exciton-hole interaction $V(\bq)$ can be taken as a contact potential, i.e. $V(\bq) = V$.  Furthermore, layer A is biased by the potential energy $\Delta$. Due to the finite tunneling term, one can diagonalize the one-particle Hamiltonian of the hole. The resulting eigenstates are called closed and open channels with energies $\epsilon_c(\bk) = \epsilon_h(\bk) + \Delta/2 + \sqrt{\frac{\Delta^2}{4} + t^2}$ and $\epsilon_o(\bk) = \epsilon_h(\bk) + \Delta/2 - \sqrt{\frac{\Delta^2}{4} + t^2}$. For later convenience, we write down  the corresponding transformation between two hole bases as
\begin{align}
& \begin{bmatrix}
a_\bk \\ h_\bk   
\end{bmatrix} 
= \begin{bmatrix}
 \frac{x_c}{\sqrt{1 + x_c^2}} &  \frac{x_o}{\sqrt{1 + x_o^2}} \\  \frac{1}{\sqrt{1 + x_c^2}} & \frac{1}{\sqrt{1 + x_o^2}}
\end{bmatrix}
\begin{bmatrix}
 h_{c \bk} \\ h_{o \bk}   
\end{bmatrix}
 \equiv U \begin{bmatrix}
h_{c \bk} \\ h_{o \bk}     
\end{bmatrix} \nonumber \\
& \equiv h(\bk) = U \tilde{h}(\bk) \text{ with}, \nonumber \\
& x_o = \frac{\Delta}{2t}\bigg(1 - \sqrt{1 + \frac{4t^2}{\Delta^2}} \bigg) \nonumber \\
& x_c = \frac{\Delta}{2t}\bigg(1 + \sqrt{1 + \frac{4t^2}{\Delta^2}} \bigg). 
\end{align}
 Here $h_{c \bk}$ and $h_{o \bk}$ are the annihilation operators for the closed and open hole channels. Moreover, the columns of $U$ are the corresponding eigenstates. 
 
 We proceed  by writing down the ${\mathcal T}$-matrix between an exciton and hole. As we have two hole channels, the scattering matrix is a 2-by-2 matrix. It is straightforward
 to show that in the open and closed channel basis (tilde basis), it is
 \begin{align}
 \label{Tmat1}
& \tilde{\btt }(\bk,ik_n) =  \begin{bmatrix}
    T_{oo}(\bk,ik_n) & T_{oc}(\bk,ik_n) \\ T_{co}(\bk,ik_n) & T_{cc}(\bk,ik_n)
\end{bmatrix} = \frac{\tilde{\bvv}}{1 - \tilde{\bvv} \tilde{\bpi}(\bk,ik_n)},     
 \end{align}
where the interaction matrix is $\tilde{\bvv} = U^\dag \bvv U$ with $V_{ij} = V\delta_{i,j}\delta_{i,1}$ and the pair propagator $\tilde{\bpi} = U^\dag \bpi U$. The un-tilded $\bvv$ and $\bpi$ are the interaction and pair propagator matrices in the original hole basis. From Eq.~\eqref{Tmat1}, one can  show that $\tilde{\btt}(\bk,ik_n) = U^\dag \btt(\bk,ik_n) U$ with $\btt$ being the ${\mathcal T}$-matrix in the original basis such that
\begin{align}
& \btt(\bk,ik_n) =  \begin{bmatrix}
    \mathcal{T}(\bk,ik_n) & 0 \\ 0 & 0
\end{bmatrix}  \\
&\mathcal{T}(\bk,ik_n) = \frac{V}{1 - V \Pi(\bk,ik_n)}  = \frac{1}{ \Pi^{2B}(0,\epsilon_T) - \Pi(\bk,ik_n)} \label{T11},  
\end{align}
where in the last step the contact interaction $V$ is eliminated in the favor of the trion binding energy $\epsilon_T<0$ and $\Pi^{2B}_{11}$ is the vacuum pair propagator. Explicitly,
\begin{align}
& \Pi(0,z) = \frac{1}{A}\sum_\bq \frac{1}{z + \Delta - q^2/2m_\mu }  \\
& \Pi^{2B}(0,\epsilon_T) = \frac{1}{A}\sum_\bq \frac{1}{\epsilon_T + i\nu - q^2/2m_\mu } 
\end{align}
where $m_\mu$ is the reduced mass of the exciton and hole. Trion bound state gives a rise to a pole for the ${\mathcal T}$-matrix. This can be seen in Fig.~\ref{two_layer}(b) where the zero-momentum spectral function of $\mathcal{T}$ is plotted as a function of frequency $\omega$. In addition to the pole at $\omega = \epsilon_T + \Delta$, the spectral function also features the usual scattering continuum at higher energies starting at $\omega = \Delta$. In the following, we discard the existence of the continuum as it appears in a non-relevant energy regime such that the ${\mathcal T}$-matrix can be written to as a pole expansion $\mathcal{T}(\bk,z) = \frac{Z}{z - (\epsilon_T + \Delta + k^2/2m_T}$, where $m_T$ is the trion mass. In our calculations, we find that the residue $Z$ is roughly momentum-independent such that $Z \approx 19 \frac{\hbar^2 |\epsilon_T|}{m_e}$, and correspondingly this is the value we used in all our computations.

\section{Derivation of the exciton self-energies}
In this section, we derive the expressions for the exciton self-energy terms $\Sigma_1$ and $\Sigma_2$. We assume $t/\Delta \ll 1$, which is likely the experimental regime of interest as $t$ is suppressed due to the spacing layer between the TMD monolayer and the
\moire system. The hopping is desired to be small also because it otherwise would lead to 
hybridization between the \moire system and the TMD monolayer, and thus to the formation of one big three-layer \moire system. As we want to use the exciton in the TMD monolayer to probe the \moire system, this is not a desired scenario.

\subsection{First order exciton self-energy $\Sigma_1$}\label{AppsecSigma1}
\begin{figure}
  \centering
    \includegraphics[width=0.8\columnwidth]{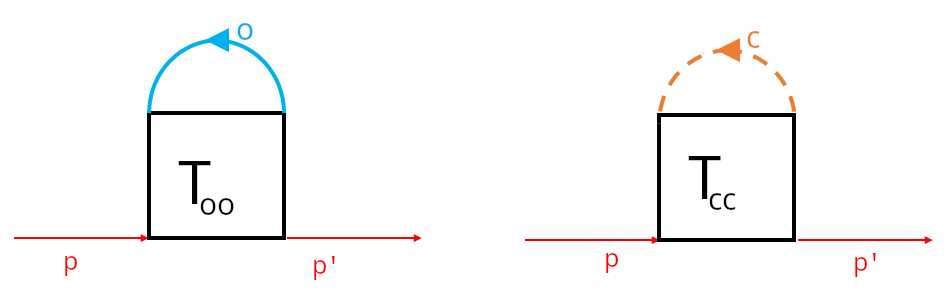}
    \caption{Exciton self-energy diagrams to first order of the scattering matrix. Red lines depict exciton propagators and are not included in the self-energy. Blue solid (orange dashed) line correspond to the hole propagators in the closed (open) channel.}
   \label{Fig2App}
\end{figure}

The first order exciton self-energy $\Sigma_1$ in terms of open and closed channels is diagrammatically depicted in Fig.~\ref{Fig2App}. Note that we have not fixed in-coming and out-going exciton momenta to be the same. Usually, they are the same but in our case,
the hole in the \moire system feels the \moire potential which can lead to finite momentum kicks by the amount of \moire reciprocal lattice vectors $\bgg_\alpha$. We therefore retain different incoming and outgoing exciton momenta, $\bp$ and $\bp'$, for now  and show below that $\bp' = \bp + \bgg_\alpha$.

As schematically presented in Fig.~\ref{Fig2App}, $\Sigma_1$ consists of two terms corresponding to the exciton interacting with the closed and open channel holes, respectively. In the following, we ignore the closed channel contribution as we take it to be approximately empty. We therefore write  $\Sigma_1$ for imaginary time  as
\begin{align}
\label{se1}
&\Sigma_1(\bp,\bp',\tau_f-\tau_i) = \frac{1}{A} \sum_{\bq \bq'} T_{oo}(\bp + \bq, \tau_f - \tau_i) \nonumber \\
& \times\langle -T h_o (\bq,\tau_i) h_o^\dag (\bq', \tau_f) \rangle \delta_{\bp',\bp + \bq - \bq'},
\end{align}
where the Green's function for the open channel is $G_{oo} (\bq,\bq',\tau_i-\tau_f) = \langle -T h_{o\bq} (\tau_i) h_{o\bq'}^\dag (\tau_f) \rangle$. In the limit of small $t/\Delta$ we obtain, up to the second order, the approximations $x_o \approx -\frac{t}{\Delta}$ and $x_c \approx \frac{\Delta}{t}\Big(1 + \frac{t^2}{\Delta^2} \Big)$. By using these, we get $|U_{12}|^2 \approx \frac{t^2}{\Delta^2}$, $|U_{22}|^2 \approx 1 - (t/\Delta)^2$, $U_{12}^* U_{22} \approx -t/\Delta$. For the self-energy Eq.~\eqref{se1} we then obtain to the leading order of $t^2/\Delta^2$
\begin{align}
\label{se2}
&\Sigma_1(\bp,\bp',\tau_f-\tau_i) \approx \frac{1}{A} \sum_{\bq \bq'} \frac{t^2}{\Delta^2} \mathcal{T}(\bp + \bq, \tau_f - \tau_i) \nonumber \\
& \times G(\bq,\bq',\tau_i - \tau_f) \delta_{\bp',\bp + \bq - \bq'},
\end{align}
where $G(\bq,\bq',\tau_i - \tau_f) = - \langle T  h_{\bq \downarrow}(\tau_i) h^\dag_{\bq' \downarrow} (\tau_f) \rangle$ is the Green's function for the holes in the MoSe$_2$ layer of the \moire system. Here we have re-introduced the spin index for the holes of the \moire system. From Eq.~\eqref{se2} we see that $\Sigma_1$ can be expressed in the limit of small $t^2/\Delta^2$ in terms of the Green's function of holes of the \moire system and the original exciton-hole ${\mathcal T}$-matrix. The tunneling between the \moire system and the exciton layer is taken into account with the multiplicative factor $t^2/\Delta^2$.

We project now the hole operators $h_{\bk \downarrow}$ to the highest \moire valence band by writing
\begin{align}
\label{projection}
&h^\dag_{ -\bk-\bgg_\alpha \sigma} = \sum_n [U_{\bk\sigma}]_{\alpha n} \gamma_{n\bk\sigma} \nonumber \\
&\approx [U_{\bk\sigma}]_{\alpha 1} \gamma_{1\bk\sigma} \equiv u_\sigma(\bk+\bgg_\alpha) \gamma_{1\bk\sigma}, 
\end{align} 
with $ \bk \in \text{mBZ}$. By furthermore transforming to  real space via the Fourier transformation $\gamma_{1\bk\sigma} = \frac{1}{\sqrt{N}} \sum_i e^{-i \bk \cdot \br_i} h^\dag_{i\sigma}$ with $N$ being the number of \moire sites $\br_i$, we obtain 
\begin{widetext}
\begin{align}
\label{se3}
&\Sigma_1(\bp,\bp',\tau) = \frac{t^2}{A\Delta^2}  \sum_{\bq \bq' \in \text{mBZ}} \sum_{\alpha \alpha'} \mathcal{T}(\bp - \bq -\bgg_\alpha, \tau) u^*_\downarrow(\bq + \bgg_\alpha) u_\downarrow(\bq' + \bgg_{\alpha'}) \sum_{ij} \frac{e^{i\bq \cdot \br_i -i\bq' \cdot \br_j}}{N} \langle - T h_{i\downarrow}(0) h_{j\downarrow}^\dag (\tau) \rangle \delta_{\bp',\bp - \bq + \bq' - \bgg_\alpha + \bgg_{\alpha'}}.
\end{align}
We see that the Green's function inside the integrand of Eq.~\eqref{se3} describes low-energy processes as $\langle - T h_{i-}(0) h_{j-}^\dag (\tau) \rangle = \langle h_{j-}^\dag (\tau) h_{i-}(0)  \rangle$. This corresponds tp the annihilation of a \moire hole instead of creating one (which would cost an energy of the on-site repulsion  $U_0$). Therefore, in principle, Eq.~\eqref{se3} should be solved such that the \moire hole operators $h_i$ are expressed with the spinon and holon operators, in a similar manner as in Ref.~\cite{Kraats2022}. This yields diagrams involving spin and holon propagators and a spin-holon bubble. Evaluating this bubble is outside the scope of the present work and we therefore use a simplified route, i.e. we assume we are far enough from the Feshbach resonance condition $\Delta = -\epsilon_T$ so that we can take the static limit of the ${\mathcal T}$-matrix, i.e. we write $\mathcal{T}(\bk,\tau) \approx  \mathcal{T}(\bk) \delta(\tau)$ with $\mathcal{T}(\bk) \equiv \mathcal{T}(\bk,\omega=0)$. From Eq.~\eqref{se3} we then get
\begin{align}
&\Sigma_1(\bp,\bp',\tau) \approx \frac{1}{A}  \frac{t^2}{\Delta^2} \delta(\tau) \sum_{\bq \bq' \in \text{mBZ}} \sum_{\alpha \alpha'} \mathcal{T}(\bp - \bq -\bgg_\alpha) u^*_\downarrow(\bq + \bgg_\alpha) u_\downarrow(\bq' + \bgg_{\alpha'}) \frac{1}{N} \sum_{ij} e^{i\bq \cdot \br_i} e^{-i\bq' \cdot \br_j} \langle  h_{j\downarrow}^\dag   h_{i\downarrow}\rangle \delta_{\bp',\bp - \bq + \bq' - \bgg_\alpha + \bgg_{\alpha'}}.
\end{align}
For the half-filled triangular \moire lattice we have $\langle h_{j\downarrow}^\dag   h_{i\downarrow} \rangle = \delta_{i,j}(\frac{1}{2}  - \langle S^z_i \rangle)$. For the in-plane magnetism or spin liquids we have $\langle S^z_i \rangle =0$ so we discard this term. Note that for the out-of-plane orders this term would be non-zero, therefore making it possible to detect out-of-plane magnetic order with $\Sigma_1$, consistent with a recent mean-field proposal of Ref.~\cite{Salvador2022}. For $\langle S^z_i \rangle =0$, we however obtain 
\begin{align}
\Sigma_1(\bp,\bp',\tau) &\approx \frac{1}{2A}  \frac{t^2}{\Delta^2} \delta(\tau) \sum_{\bq \bq' \in \text{mBZ}} \sum_{\alpha \alpha'} \mathcal{T}(\bqq - \bq -\bgg_\alpha) u^*_\downarrow(\bq + \bgg_\alpha) u_\downarrow(\bq' + \bgg_{\alpha'}) \delta_{\bq,\bq'} \delta_{\bp',\bp - \bq + \bq' - \bgg_\alpha + \bgg_{\alpha'}}  \nonumber \\
& = \frac{1}{2A}  \frac{t^2}{\Delta^2} \delta(\tau) \sum_{\bk} \sum_{\lambda} \mathcal{T}(\bp - \bk) u^*_\downarrow(\bk) u_\downarrow(\bk + \bgg_{\lambda}) \delta_{\bp',\bp + \bgg_\lambda}
\end{align}
\end{widetext}
where in the last line we have carried out a change of integration variables as $\bk = \bq + \bgg_\alpha$ and $\bgg_\lambda = \bgg_{\alpha'} - \bgg_\alpha$ such that the sum over $\bk$ is taken over the full momentum space, not just the \moire BZ. Finite terms for $\Sigma_1$ are therefore
\begin{align}
\label{Umklapp}
& \Sigma_1(\bp,\bp + \bgg_\lambda,\tau) =  \frac{1}{2A}\frac{t^2}{\Delta^2} \delta(\tau) \sum_{\bk} \mathcal{T}(\bp - \bk) u^*_\downarrow(\bk) u_\downarrow(\bk + \bgg_{\lambda}) 
\end{align}
which is the Umklapp term  discussed in the main text, i.e. the first term of Eq.~\eqref{Selfenergy1}. 
As noted in the main text, for the in-plane magnetism or spin liquid, this Umklapp term is not enough to obtain information about the in-plane magnetic correlations within the system. 

To get the estimate for the zero-momentum energy shift due to the Umklapp potential, i.e. $\delta E_{\text{Um}}$, we consider Eq.~\eqref{Umklapp} with $\bp =0$ and $\bgg_\lambda = 0$, such that  we have 
\begin{align}
\delta E_{\text{Um}} = \frac{t^2}{2A \Delta^2} \sum_\bk \mathcal{T}(\bk) |u_\downarrow(\bk)|^2 = \frac{t^2 Z}{\Delta^2 a_m^2} \beta,
\end{align}
where $\beta$ is the $\bk$-integral in dimensionless units, i.e. $\beta = \frac{1}{8\pi^2} \int d \tilde \bk \frac{|u_\downarrow(\tilde \bk)|^2}{1 - \tilde \Delta - \alpha_r \tilde k^2}$ with $\alpha_r = 1/(|\epsilon_T| 2 m_T a_m^2)$ and tilded variables are dimensionless. As $\beta \sim \mathcal{O}(1)$, we have $\delta E_{\text{Um}} \sim \mathcal{O}(\frac{t^2 Z}{\Delta^2 a_m^2} )$.

\begin{figure}
  \centering
    \includegraphics[width=0.5\textwidth]{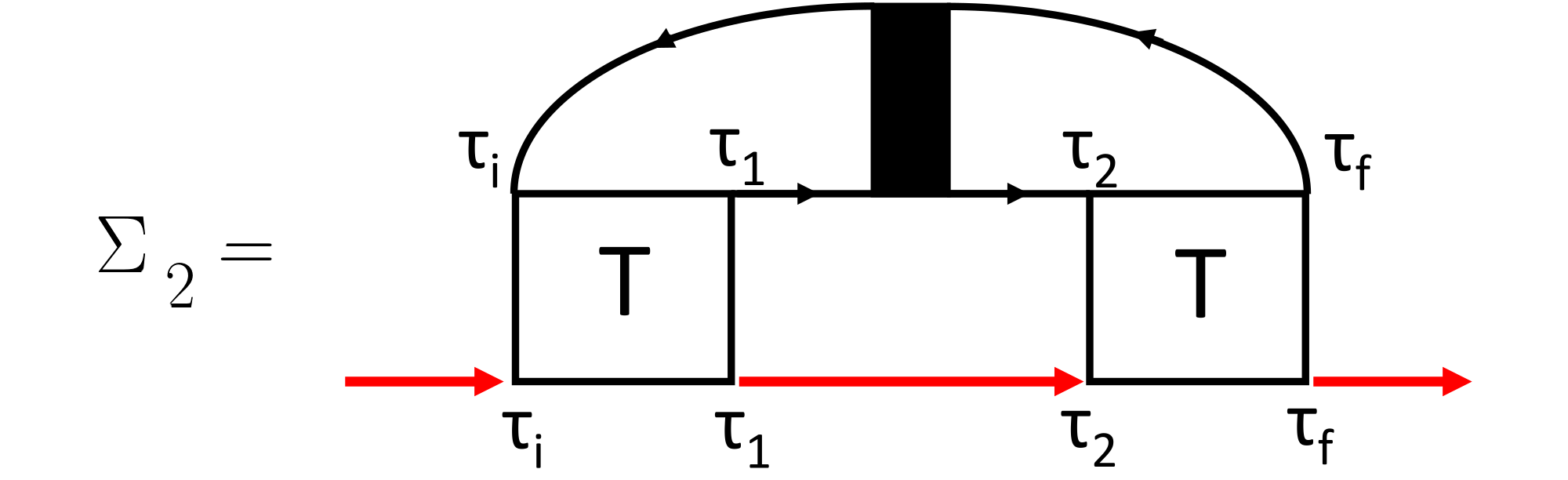}
    \caption{The second order exciton self-energy $\Sigma_2$ in terms of the ${\mathcal T}$-matrix. Black box corresponds to the full two-particle propagator of \moire holes.}
   \label{Fig4S}
\end{figure}

\subsection{Second order exciton self-energy $\Sigma_2$ and the spin-spin correlator}\label{AppSecSigma2}
After having calculated the Umklapp contribution arising from $\Sigma_1$, we consider another type of exciton self-energy depicted in Fig.~\ref{Fig4S}, which we denote $\Sigma_2$. This self-energy is second order in the ${\mathcal T}$-matrix and contains information about the dynamical spin-spin correlations of the holes in the \moire lattice.

We can write $\Sigma_2$ in  imaginary time  to leading order in $t^4/\Delta^4$ as
\begin{widetext}
\begin{align}
\label{se2_1}
 \Sigma_2(\bp,\bp',\tau = &\tau_f - \tau_i) = \frac{t^4}{\Delta^4}\frac{1}{A^2} \sum_\bq \int_0^\beta d\tau_1 \int_0^\beta d\tau_2 \mathcal{G}(\bp + \bq, \tau_2 - \tau_1) \sum_{\bk \bk'}  \mathcal{T}(\bp+ \bk + \bq, \tau - \tau_i) \mathcal{T}(\bp + \bk' + \bq, \tau_f - \tau_2 ) \nonumber \\
& \times \langle h^\dag_{\bk'+\bq' \downarrow}(\tau_f) h_{\bk'\downarrow}(\tau_2) h_{\bk \downarrow}^\dag (\tau_1) h_{\bk+\bq \downarrow}(\tau_i) \rangle \delta_{\bp',\bp + \bq - \bq'}.
\end{align}
As in case of  $\Sigma_1$, we again assume that we are sufficiently far away from the Feshbach resonance such that we can approximate the ${\mathcal T}$-matrix with its static limit, i.e. $\mathcal{T}(\bk,\tau) \approx \mathcal{T}(\bk) \delta(\tau)$. We can then cast Eq.~\eqref{se2_1} as
\begin{align}
& \Sigma_2(\bp,\bp',\tau) \approx \frac{t^4}{\Delta^4}\frac{1}{A^2} \sum_\bq \mathcal{G}(\bp+ \bq, \tau) \sum_{\bk \bk'} \mathcal{T}(\bp+ \bk + \bq) \mathcal{T}(\bp + \bk' + \bq) \langle h^\dag_{\bk'+\bq' \downarrow}(\tau) h_{\bk' \downarrow}(\tau) h_{\bk \downarrow}^\dag h_{\bk+\bq \downarrow} \rangle \delta_{\bp',\bp + \bq - \bq'}. 
\end{align}
We proceed now the same way as before, i.e. project $h_{\bk \downarrow}$ to the highest \moire valence band with Eq.~\eqref{projection} and transform to the real space of the \moire lattice. This gives 
\begin{align}
\label{se2_mid2}
& \Sigma_2(\bp,\bp',\tau) \approx \frac{t^4}{\Delta^4}\frac{1}{A^2} \sum_{\bq \bq' \in \text{mBZ}}\sum_{\lambda \lambda'} \mathcal{G}(\bp - \bq - \bgg_\lambda,\tau) \sum_{\bk \bk' \in \text{mBZ}}\sum_{\alpha\beta} \mathcal{T}(\bp - \bk - \bgg_\alpha - \bq - \bgg_\lambda) \mathcal{T}(\bp - \bk' - \bgg_\beta - \bq -\bgg_\lambda) \nonumber \\
& \times u_\downarrow (\bk' + \bq' + \bgg_\beta + \bgg_{\lambda'}) u_\downarrow^*(\bk'+\bgg_\beta) u_\downarrow(\bk+\bgg_\alpha) u_\downarrow^*(\bk + \bgg_\alpha + \bq + \bgg_\lambda) \delta_{\bp',\bp - \bq - \bgg_\lambda + \bq' + \bgg_{\lambda'}} \nonumber \\
& \times \frac{1}{N}\sum_{i,i',i'',i'''} e^{-i ( \bq' + \bk' ) \cdot \br_i} e^{i \bk' \cdot \br_{i'}}  e^{-i \bk \cdot \br_{i''}} e^{i(\bq + \bk ) \cdot \br_{i'''}} \langle T h^\dag_{i\downarrow}(\tau) h_{i'\downarrow}(\tau) h^\dag_{i'' \downarrow} h_{i'''\downarrow} \rangle .
\end{align}
By focusing on the last line, one can see that  if $i'' \neq i'''$ or $i \neq i'$, the term $\langle T h^\dag_{i\downarrow}(\tau) h_{i'\downarrow}(\tau) h^\dag_{i'' \downarrow} h_{i'''\downarrow} \rangle$ is proportional to $ e^{\tau U_0}$. Later, when transforming to  Matsubara frequencies, this quickly oscillating term suppresses the  self-energy by
a factor of $1/U_0$. As the repulsive interaction $U_0$ is the dominant energy scale in our lattice model, we can ignore such non-local terms with $i'' \neq i'''$ or $i \neq i'$ and instead take $i'= i$ and $i''' = i'' \equiv j$. Hence, we get
\begin{align}
\label{se2_2}
& \Sigma_2(\bp,\bp',\tau) \approx \frac{t^4}{\Delta^4}\frac{1}{A^2} \sum_{\bq \bq' \in \text{mBZ}}\sum_{\lambda \lambda'} \mathcal{G}(\bp - \bq - \bgg_\lambda,\tau) \sum_{\bk' \bk \in \text{mBZ}}\sum_{\alpha\beta} \mathcal{T}(\bp - \bk - \bgg_\alpha - \bq - \bgg_\lambda) \mathcal{T}(\bp - \bk' - \bgg_\beta - \bq -\bgg_\lambda) \nonumber \\
& \times u_\downarrow (\bk' + \bq' + \bgg_\beta + \bgg_{\lambda'}) u_\downarrow^*(\bk'+\bgg_\beta) u_\downarrow(\bk+\bgg_\alpha) u_\downarrow^*(\bk + \bgg_\alpha + \bq + \bgg_\lambda) \delta_{\bp',\bp - \bq - \bgg_\lambda + \bq' + \bgg_{\lambda'}} \nonumber \\
& \times \frac{1}{N}\sum_{ij} e^{-i \bq'  \cdot \br_i}   e^{i \bq \cdot \br_j} \langle T h^\dag_{i\downarrow}(\tau) h_{i\downarrow}(\tau) h^\dag_{j \downarrow} h_{j\downarrow} \rangle.
\end{align}
This can be further simplified at  half-filling by noting that $h_{i\downarrow}^\dag h_{i\downarrow} = \frac{1}{2} - S^z_i$. By inserting this to Eq.~\eqref{se2_2} and further transferring to  Matsubara frequency space, we obtain 
\begin{align}
\label{se2_mid}
& \Sigma_2(\bp,\bp',i\omega_n) = \frac{1}{N\beta}  \frac{t^4}{\Delta^4}\frac{1}{A^2} \sum_{\bq \bq' \in \text{mBZ}}\sum_{\lambda \lambda'} \sum_{iq_n} \mathcal{G}(\bp - \bq - \bgg_\lambda, i\omega_n + iq_n) \delta_{\bp',\bp - \bq - \bgg_\lambda + \bq' + \bgg_{\lambda'}} \nonumber \\
& \times \Big[ \sum_\bk \mathcal{T}(\bp - \bk - \bq - \bgg_\lambda) U_-( \bk) U_-^*(\bk + \bq + \bgg_\lambda) \Big] \Big[ \sum_{\bk'} \mathcal{T}(\bp - \bk' - \bq - \bgg_\lambda) U^*_-( \bk') U_-(\bk' + \bq' + \bgg_{\lambda'}) \Big] \nonumber \\
&\times \int_0^\beta d\tau e^{-iq_n \tau} \langle T S^z_{\bq'}(\tau) S^z_{-\bq} \rangle .
\end{align}
 Here we have discarded reducible diagrams. One should note that  the sums of $\bk$ and $\bp$ are taken over the whole momentum space. The last line in Eq.~\eqref{se2_mid} is the spin-spin correlator. If we assume the translational invariance for the spin-spin correlator (which should hold for  a $120\degree$ AFM,  FM, or a spin liquid), i.e. that $\langle S^z_{j + \Delta j} (\tau) S^z_j \rangle$ depends only on the relative distance $\Delta j$ between two lattice sites $j$ and $ j +\Delta j$, we find $\bq' = \bq$ such that
\begin{align}
& \Sigma_2(\bp,\bp',i\omega_n) =  \frac{1}{N\beta} \frac{t^4}{\Delta^4}\frac{1}{A^2} \sum_{\bq  \in \text{mBZ}}\sum_{\lambda \lambda'} \sum_{iq_n} \mathcal{G}(\bp - \bq - \bgg_\lambda, i\omega_n + iq_n)  \chi_{zz}(\bq,-iq_n) \delta_{\bp',\bp - \bgg_\lambda + \bgg_{\lambda'} }  \nonumber \\
& \times \Big[ \sum_\bk \mathcal{T}(\bp - \bk - \bq - \bgg_\lambda) u_\downarrow( \bk) u_\downarrow^*(\bk + \bq + \bgg_\lambda) \Big] \Big[ \sum_{\bk'} \mathcal{T}(\bp - \bk' - \bq - \bgg_\lambda) u^*_\downarrow( \bk') u_\downarrow(\bk' + \bq + \bgg_{\lambda'}) \Big]
\end{align}
where the spin-spin correlator is $ \chi_{zz}(\bq,iq_n) = \int_0^\beta d\tau e^{iq_n \tau} \langle T S^z_{\bq}(\tau) S^z_{-\bq} \rangle$. 
By defining a new label as $\bgg_\alpha = \bgg_{\lambda'} - \bgg_\lambda$ we get
\begin{align}
&\Sigma_2(\bp,\bp + \bgg_\alpha,i\omega_n) = \frac{1}{N\beta}\frac{t^4}{\Delta^4}\frac{1}{A^2} \sum_{\bq  \in \text{mBZ}}\sum_{\lambda}  \sum_{iq_n} \mathcal{G}(\bp - \bq - \bgg_\lambda, i\omega_n + iq_n) \chi_{zz}(\bq,-iq_n)  \nonumber \\  
& \times \Big[ \sum_\bk \mathcal{T}(\bp - \bk - \bq - \bgg_\lambda) u_\downarrow( \bk) u_\downarrow^*(\bk + \bq + \bgg_\lambda) \Big] \Big[ \sum_{\bk'} \mathcal{T}(\bp - \bk' - \bq - \bgg_\lambda) u^*_\downarrow( \bk') u_\downarrow(\bk' + \bq + \bgg_\lambda + \bgg_\alpha) \Big].
\end{align}
We can express this as
\begin{align}
&\Sigma_2(\bp,\bp + \bgg_\alpha,i\omega_n) = \frac{1}{N\beta}\frac{t^4}{\Delta^4}\frac{1}{A^2} \sum_{\bq} \sum_{iq_n} \mathcal{G}(\bp - \bq, i\omega_n - iq_n) \chi_{zz}(\bq,iq_n)  g(\bp,-\bq) g(\bp-\bq, \bq + \bgg_\alpha) \label{final_se2}, \\
& g(\bp,\bq) =  \sum_\bk \mathcal{T}(\bp - \bk + \bq) u_\downarrow( \bk) u_\downarrow^*(\bk - \bq), \label{2nd_order_SE}
\end{align}
which is the same expression as in the main text.

We have therefore obtained the Umklapp self-energy term of Eq.~\eqref{Umklapp} and the term containing the spin-spin correlator given by Eq.~\eqref{2nd_order_SE}. The major approximation to derive these terms was to take the static limit of the exciton-hole ${\mathcal T}$-matrix. If one wanted to take into account the full frequency dependency of the 
${\mathcal T}$-matrices, evaluating the diagram of Fig.~\ref{Fig4S}  would be very challenging as it would contain a a two-particle Green's function that would split into complicated spinon-holon correlators. We leave this topic for future studies.

\section{Additional details concerning computing $\Sigma_2$ with  SCBA}\label{numerics_sec}
In this section, we provide further practical details regarding our SCBA computations to evaluate Eq.~\eqref{Selfenergy2}. To this end, it is convenient to write $\Sigma_2$ by folding the exciton dispersion to the mBZ such that $\Sigma_2$ reads 
\begin{align}
\label{Sigma2_scba}
&[\Sigma_2(\bp,ip_n)]_{nn'} = \frac{1}{\beta N}\sum_{\substack{\bq\in\text{mBZ}\\ iq_n}} \chi_{zz}(q) [\tilde{W}^\dag(\bp,\bq) \mathcal{G}(\bp+\bq,ip_n + iq_n) \tilde{W}(\bp,\bq) ]_{nn'} 
\end{align}
with
\begin{align}
& \tilde{W}_{nm}(\bp,\bq)  \equiv \frac{1}{A}\sum_{\bk,\alpha\lambda} \mathcal{T}(\bp + \bgg_\alpha - \bk)  S^*_n(\bp+ \bgg_\alpha + \bq + \bgg_\lambda)  S_{m}(\bp+ \bgg_\alpha) U_\sigma(\bk+\bq + \bgg_\lambda) U_\sigma^*(\bk),
\end{align}
where we have expressed the exciton degrees of freedom in the eigenbasis of the Umklapp potential, i.e. $X_{\bp+\bgg_\alpha} = \sum_n S_n(\bp+\bgg_\alpha) \gamma_{X n}(\bp)$.
Here, $S_n(\bp)$ are the Bloch functions of the excitons and $\gamma_{X n}(\bp)$ are the exciton annihilation operators for exciton bands $n$ and momenta $\bp$, corresponding to exciton energies $\epsilon_{X n}(\bp)$. Moreover, the excitonic Green's function $\mathcal{G}(Q)$ in the new basis reads $[\mathcal{G}(\bp,\tau)]_{nm} = -\langle T \gamma_{X n}(\bp,\tau) \gamma_{X m}^\dag(\bp) \rangle$ such that $\mathcal{G}^{-1}_X(Q) = \mathcal{G}_0^{-1}(p) - \Sigma_2(p)$ and $[\mathcal{G}_0^{-1}(p)]_{nm} = \delta_{nm}[ip_n - \epsilon_{X n}(\bp)]$ with the notation $p = (\bp,ip_n)$. 

By using the LSWT result for the spin-spin correlator $\chi_{zz}$ and the fact that we are considering a single exciton, we can carry out the Matsubara sum in
Eq.~\eqref{Sigma2_scba} giving at zero temperature 
\begin{align}
\label{SCBA}
&[\Sigma_2(\bp,\omega)]_{nn'} = \frac{1}{N} \sum_{\bq \in \text{mBZ}} ( 2u_{\bq} v_{\bq} - ( u_\bq^2 + v_\bq^2)) [\tilde{W}^\dag(\bp,\bq) \mathcal{G} (\bp+\bq, \omega + 0^+ -\epsilon_\bq) \tilde{W}(\bp,\bq) ]_{nn'}.   
\end{align}
As the exciton Green's function $\mathcal{G}$ at the right hand side of Eq.~\eqref{SCBA} depends on the self-energy $\Sigma_2$, Eq.~\eqref{SCBA} needs to be solved self-consistently. To this end, we used an initial ansatz $\mathcal{G} =\mathcal{G}_0$ and solved Eq.~\eqref{SCBA} iteratively by computing a new value for $\Sigma_2$ with Eq.~\eqref{SCBA} which was then used to obtain a new $\mathcal{G}$ via $\mathcal{G}^{-1} = \mathcal{G}_0^{-1} - \Sigma_2$. This procedure was repeated untill $\Sigma_2$ converged to a stable solution. 
Note that the second order perturbation results of Fig.~\ref{FigRes3} are obtained by terminating this loop after the first iteration, see Sec.~\ref{2nd_ord_app}. 
As the term arising from the coherence factors of the spin wave excitations is strongly peaked around the Dirac points in case of the AFM, we used non-uniform  grid with more momentum points around the Dirac points to evaluate the momentum integral of Eq.~\eqref{SCBA}.

We see from Eq.~\eqref{SCBA} that in general the exciton self-energy and the Green's function $\mathcal{G}$ are matrices expressed in the exciton band basis. This representation arises when we folded the dispersion of the free exciton to the \moire BZ. This is an alternative presentation to Eq.~\eqref{final_se2} where the exciton self-energy and the Green's function are expressed in terms of the \moire reciprocal vectors $\bgg_\alpha$. The expressions in the main text are for simplicity presented in terms of $\bgg_\alpha$ and the results of the main text are obtained from Eq.~\eqref{SCBA} by considering only the lowest exciton band, i.e. $n=1$. We have numerically checked that by including the second band has a negligible effect on our results.
\end{widetext}

\begin{figure}[ht!]
\centering
\includegraphics[width=0.9\columnwidth]{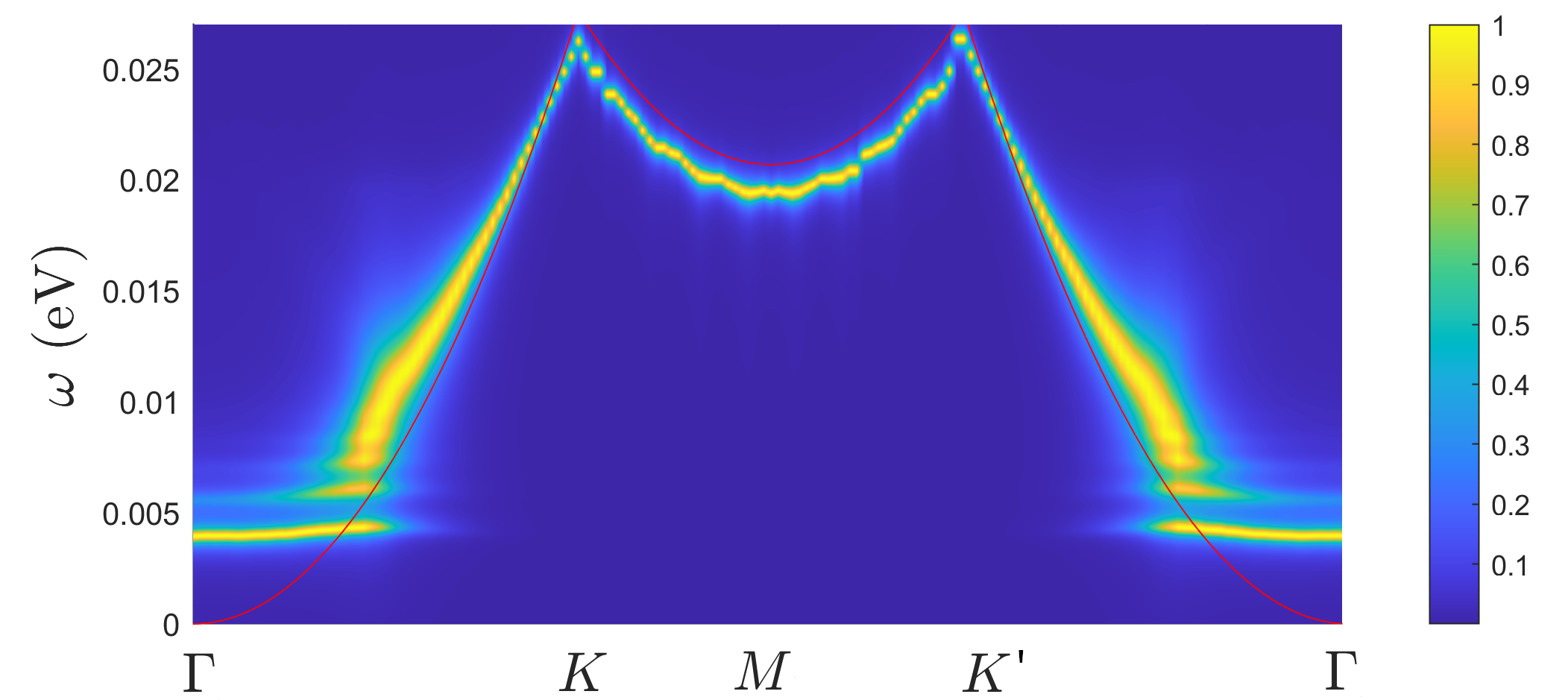}
\caption{Exciton spectral function $A(\bp,\omega)$ for the twist angle  $\theta=3^\circ$  and an in-plane AFM along the $\Gamma-K-K'-\Gamma$ path within the mBZ as a function of the momentum and energy. The red solid line gives the bare exciton dispersion. 
}
\label{FigGKKG}
\end{figure}

In the main text, numerical results are shown along the $\Gamma$-$M$ path in the mBZ. For completeness, in Fig.~\ref{FigGKKG} the spectral function is shown along the $\Gamma-K-K'-\Gamma$ path. 
We see that the spectral features are  similar those in Fig.~\ref{FigRes1}, i.e.\ there is a clear low energy  quasiparticle peak from the exciton-polaron state and broader peaks at 
higher energies corresponding to geometric string excitations.

\section{Physical picture for the presence of string excitations}\label{string_ex_sec}
In this section, we give elaborate on the physical reason  why the string excitations are present in case of the AFM  but not for the FM. We start by noting that after approximating the exciton-hole ${\mathcal T}$-matrix with its static limit, the effective exciton-hole interaction Hamiltonian for the lowest exciton band reads
\begin{align}
\label{AppSE_e1}
H_{X-h} = \sum_{\bp \bq \in \{ \text{mBZ} \}} V(\bp,\bq) \gamma_{\bp  + \bq}^\dag \gamma_\bp s_\bq^z,
\end{align}
with
\begin{align}
  V(\bp,\bq) = \frac{1}{\sqrt{N}A} \tilde{W}_{11}(\bp,\bq)    
\label{Veqn}
\end{align}
where $\gamma_\bp \equiv \gamma_{X1}(\bp)$ is the exciton annihilation operator for the lowest exciton band (See the previous section for the notation). In other words, the exciton can scatter by creating a spin flip in the \moire system. This is due to the fact that for the in-plane magnetic order, $s^z_i$  flips the spin of \moire site $i$. By Fourier transforming Eq.~\eqref{AppSE_e1} to  real space of \moire sites $h_i$ and exciton sites $\gamma_i = \frac{1}{\sqrt{N}} \sum_\bp e^{i \bp \cdot \br_i} \gamma_\bp$, one finds
\begin{align}
\label{AppSE_e2}
H_{X-h} =  \sum_{i,i',j'} V(\br_{i'} - \br_{j'},\br_{i'} - \br_i) \gamma_{i'}^\dag \gamma_{j'} s^z_i.
\end{align}
We therefore see explicitly that the exciton hopping can create spin flips, analogous to a hole  hopping in a AFM lattice.  We have numerically checked that $V(\br_1,\br_2)$ is largest for the local processes, i.e. $\br_{i'} = \br_{j'} = \br_i$, such that the exciton mainly creates  spin flips at the nearest \moire site. It follows that when 
the exciton moves around, it  creates spin flips in its wake. In other words, spin flips take place in the proximity of the exciton and hence the exciton can become dressed due to the cloud of spin flips around it.

\begin{figure}[ht!]
  \centering
    \includegraphics[width=1.0\columnwidth]{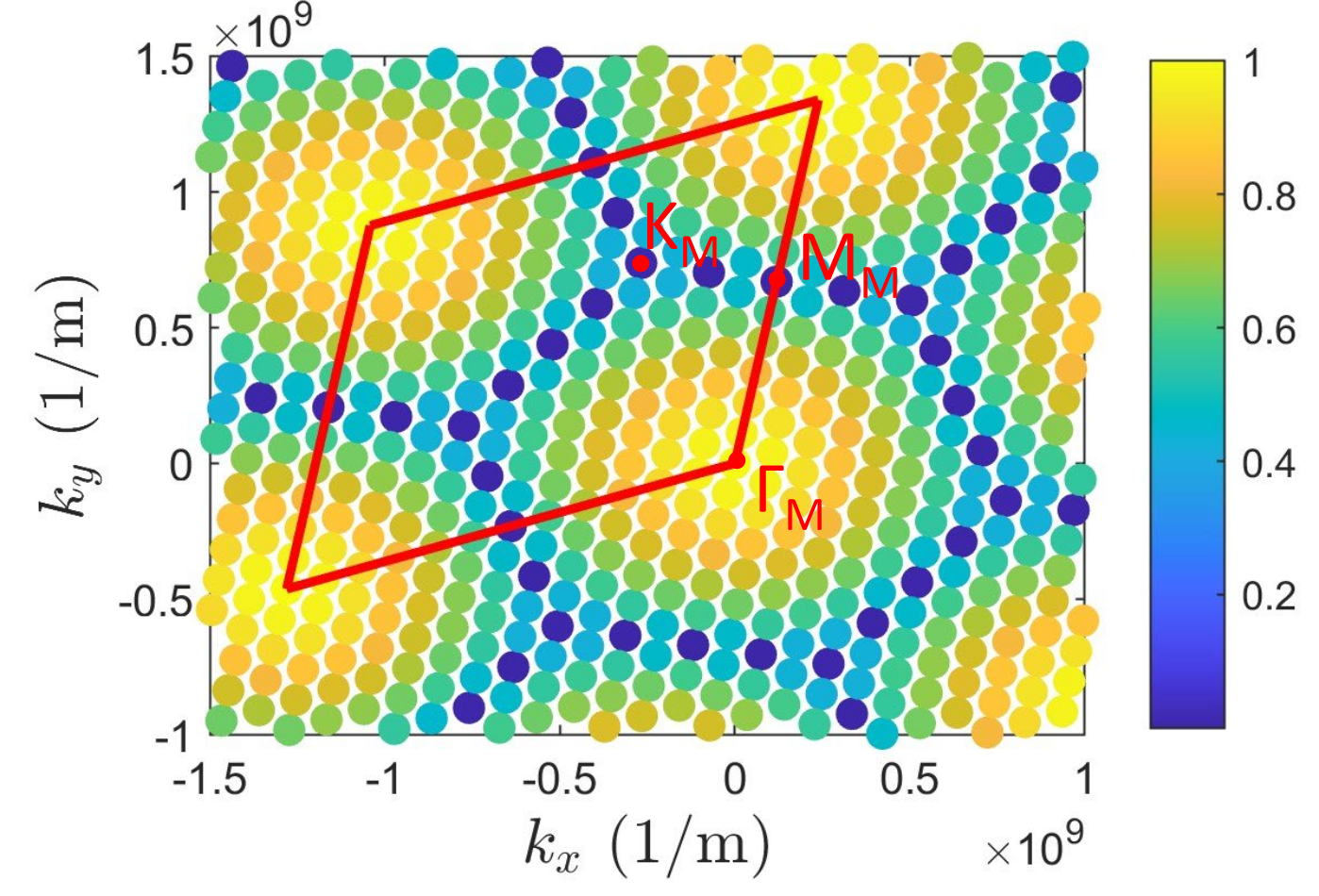}
    \caption{$V(0,\bq)/\max[V(0,\bq]$ in Eq.~\eqref{Veqn} as a function of momentum $\bq$ for $\theta = 2.6^\circ$. Red parallelogram indicates the momentum space unit cell.}
   \label{Fig_S}
\end{figure}

We see that in Eq.~\eqref{AppSE_e2} there exists also non-local hopping terms for the exciton. These are not included in the heuristic picture acquired from 
Eq.~\eqref{Tmatrix} and discussed in Sec.~\ref{phys_sec}. Indeed, we have numerically checked that such hopping terms given by Eq.~\eqref{AppSE_e2} are much smaller in amplitude compared to the dominant local term. The reason for the existence of non-local terms  lies in the fact that Eq.~\eqref{AppSE_e2} is obtained by projecting the hole (exciton) operators to the highest \moire valence band (lowest exciton band). Such a projection scheme yields the interaction vertex $V(\bp,\bq)$ of Eq.~\eqref{AppSE_e1} that depends on the momenta of both exciton and hole. This is in contrast to the original exciton-hole ${\mathcal T}$-matrix of Eq.~\eqref{Tmatrix} that depends only on the total center-of-mass momentum.

It is now illuminating to consider again the momentum-space representation of Eq.~\eqref{AppSE_e1}. As can be seen in Fig.~\ref{FigRes1}, string excitations take place for small exciton momentum $\bp$. Motivated by this observation, we consider $\bp=0$ and correspondingly in Fig.~\ref{Fig_S}  show $V(0,\bq)$ as a function of the spin wave momentum $\bq$. From Fig.~\ref{Fig_S} we see that the creation of spin wave excitations is most prominent for small momenta $\bq$. Therefore, string excitations are present (absent) for the AFM (FM) order as local spin flips, that preserve their shape and compactness, can (cannot) be constructed from the linearly (quadratically) dispersing spin wave excitations near $\bq=0$.

\begin{figure}[ht!]
\centering
\includegraphics[width=0.9\columnwidth]{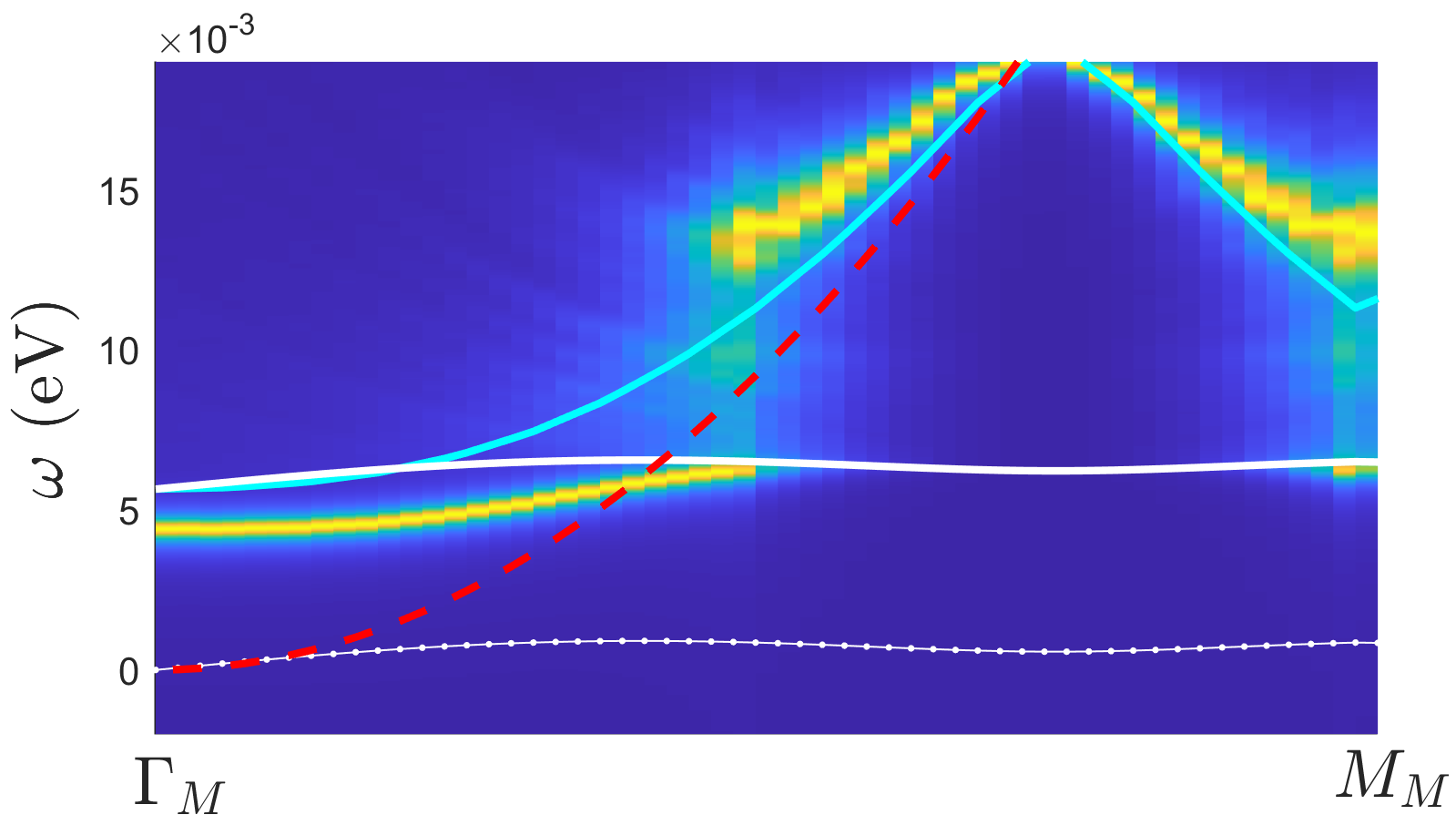}
\caption{Exciton spectral weight $A(\bp,\omega)$ at $\theta =3^\circ$ as a function of the frequency and the momentum obtained with the non-self-consistent second order perturbation theory. Red dashed, solid cyan, das-dotted white and solid white lines correspond, respectively, to the bare exciton dispersion, exciton dispersion in the presence of the Umklapp potential $\Sigma_1$, spin wave spectrum and $\omega_c$ (see text for meaning of $\omega_c$). These results should be compared to the corresponding SCBA results of Fig.~\ref{FigRes1}. We see that the string excitations are absent, highlighting the importance of our SCBA approach.}
\label{FigRes3}
\end{figure}
\section{Second order perturbation theory}\label{2nd_ord_app}
To demonstrate that the non-perturbative SCBA is required to reveal the string excitations, in Fig.~\ref{FigRes3} we show the exciton spectral weight for  $\theta =3^\circ$ as a function of momentum and frequency obtained with the non-self-consistent second  order perturbation theory by excluding the contribution of $\Sigma_2$ within the integrand of Eq.~\eqref{Selfenergy2}, i.e. by using the non-interacting exciton Green's function
\begin{align}
\label{2nd_or_non_sc}    
&{\mathcal G}^{-1}({\mathbf p}',{\mathbf p},i\omega_n)=i\omega_n-\epsilon^x_{\bp}-\Sigma_1({\mathbf p}',{\mathbf p})
\end{align}
in Eq.~\eqref{Selfenergy2}. We see that the string excitations are absent in Fig.~\ref{FigRes3} as compared to the SCBA results shown in Fig.~\ref{FigRes1}, underscoring the importance of the non-perturbative SCBA to describe the spectral properties of the probe exciton correctly.

Both the second order results of Fig.~\ref{FigRes3} and the results of the SCBA computations shown in Fig.~\ref{FigRes1} reveal that a well-defined quasi-particle exists only 
for small momenta,  and that it  is strongly  damped elsewhere in the mBZ, although they disagree concerning its energy. To explain the damping within the second order perturation theory, we consider the integrand of Eq.~\eqref{Selfenergy1} with Eq.~\eqref{2nd_or_non_sc} and $\bp' = \bp$ and note that $\epsilon^x_\bp + \Sigma_1(\bp,\bp) \sim \delta E_{\text{Um}} + p^2/(2 m_{\text{Um}})$. Here $\delta E_U$ ($m_{\text{Um}}$) is the zero-momentum energy shift (effective mass) of the exciton due to the Umklapp potential. From Eq.~\eqref{Selfenergy2} we can deduce that   the particle-hole scattering continuum should for a given momentum $\bp$ start at the critical frequency $\omega_c(\bp) = \min_\bq [ \omega_\bq + \delta E_{\text{Um}} + (\bp - \bq)^2/(2 m_{\text{Um}})]$. As the spin wave dispersion $\omega_\bq$ remains rather flat compared to the exciton energies, for large enough $\bp$ this condition yields $\omega_c(\bp) = \delta E_{\text{Um}} + \omega_\bp$. In Fig.~\ref{FigRes3} we have plotted this estimate as a white line. We see that once the energy of the well-defined quasi-particle mode crosses this condition, it becomes damped. Hence, the condition for $\omega_c$ explains the damping of the exciton-polaron mode. Similar damping and the transfer of the spectral weight from the quasi-particle mode to the excited states and broad continuum are also seen in case of the strong coupling
SCBA results of Fig.~\ref{FigRes1}.

\bibliography{bib_file}
\end{document}